**Identification of key genes related to the mechanism and prognosis of lung squamous cell carcinoma using bioinformatics analysis**

Miaomiao Gao[a,#], Weikaixin Kong[b,#], Zhuo Huang[b,*] and Zhengwei Xie[a,*]

[#]These authors contributed equally to this work

[a]*Department of Pharmacology and Department of Biomedical Informatics, School of Basic Medical Sciences, Peking University Health Science Center, 38 Xueyuan Lu, Haidian district, Beijing 100191, China*

[b]*Department of Molecular and Cellular Pharmacology, School of Pharmaceutical Sciences, Peking University Health Science Center, 38 Xueyuan Lu, Haidian district, Beijing 100191, China*

* To whom correspondence should be addressed.

ZhengweiXie

Tel: 86-10-82802798; Fax: 86-10-82802798; Email: xiezhengwei@hsc.pku.edu.cn.

Zhuo Huang

E-mail: huangz@hsc.pku.edu.cn

**Abstract**

Objectives

Lung squamous cell carcinoma (LUSC) often diagnosed as advanced with poor prognosis. The mechanisms of its pathogenesis and prognosis require urgent elucidation. This study was performed to screen potential biomarkers related to the occurrence, development and prognosis of LUSC to reveal unknown physiological and pathological processes.

Materials and Methods

Using bioinformatics analysis, the lung squamous cell carcinoma microarray datasets from the GEO and TCGA databases were analyzed to identify differentially expressed genes (DEGs). Furthermore, PPI and WGCNA network analysis were integrated to identify the key genes closely related to the process of LUSC development. In addition, survival analysis was performed to achieve a prognostic model that accomplished a high level of prediction accuracy.

Results and Conclusion

Eighty-five up-regulated and 39 down-regulated genes were identified, on which functional and pathway enrichment analysis was conducted. GO analysis demonstrated that up-regulated genes were principally enriched in epidermal development and DNA unwinding in DNA replication. Down-regulated genes were mainly involved in cell adhesion, signal transduction and positive regulation of inflammatory response. After PPI and WGCNA network analysis, eight genes, including AURKA, RAD51, TTK, AURKB, CCNA2, TPX2, KPNA2 and KIF23, have been found to play a vital role in LUSC development. The prognostic model contained 20 genes, 18 of which were detrimental to prognosis. The AUC of the established prognostic model for predicting the survival of patients at 1, 3, and 5 years was 0.828, 0.826 and 0.824, respectively. To conclude, this study identified a number of biomarkers of significant interest for additional investigation of the therapies and methods of prognosis of lung squamous cell carcinoma.

## 1. Introduction

Lung cancer is among the deadliest malignancies globally and can be categorized into two main types: small cell lung cancer (SCLC) and non-small cell lung cancer (NSCLC). NSCLC, accounting for approximately 85% of lung cancer cases, is principally divided into lung adenocarcinoma (LUAD) and lung squamous cell carcinoma (LUSC) depending on pathogenesis and histological morphology [1-3]. Many therapeutic methods are currently being used to treat lung cancer, such as surgical resection, chemotherapy, radiotherapy and targeted therapy. Because NSCLC is insensitive to radiotherapy and resistant to anticancer drugs, targeted therapy has become the most effective way to treat patients whose tumors cannot be surgically removed [4,5]. In recent years, targeted therapies have undergone considerable development, and some effective molecular targets have been identified, such as epidermal growth factor receptor (EGFR), anaplastic lymphoma kinase (ALK), *etc*. It has been confirmed that these targets are successful in lung adenocarcinoma, but not in lung squamous cell carcinoma, because the two major subtypes have different mutation profiles [6-8]. In order to better diagnose and treat patients with LUSC, an investigation of novel biomarkers is required. LUSC, which accounts for 30% of cases of NSCLC, is more common in middle-aged and older men and has a high rate of metastasis and recurrence [9]. NSCLC patients are mostly diagnosed during the advanced stage, their 5-year survival rate lower than that of early stage patients [4,10,11]. Thus, further study of prognostic markers is required so as to personalize cancer treatment. Although a number of studies have reported on LUSC-related genes and prognostic markers, the specific molecular mechanisms in the pathogenesis and progression of LUSC have not been systematically evaluated, which constrains the potential for early diagnosis and treatment [12,13]. An in-depth understanding of the molecular mechanisms involved in the occurrence and development of lung cancer may provide a more effective strategy for early detection and subsequent clinical treatment.

Therefore, novel promising biomarkers or potential drug treatments are urgently required.

Genomic microarrays and high-throughput sequencing technology combined with bioinformatics analysis has gradually become a powerful tool for the discovery of disease biomarkers and related pathways. In the present study, three RNA microarray datasets from the GEO database were analyzed to identify differentially expressed genes (DEGs) in lung squamous cell carcinoma compared with normal tissues. Gene ontology (GO) and Kyoto Encyclopedia of Genes (KEGG) analyses were performed to identify key genes and pathways which possibly influence the pathogenesis of LUSC. On the other hand, RNAseq data from the TCGA database were also used to screen out DEGs for weighted gene co-expression network analysis (WGCNA) so as to obtain hub genes associated with the occurrence and development of LUSC. Eventually, key genes at the intersection of these two databases identified genes that are critical to the disease. In addition, we utilized patient clinical information from DEGs in the TCGA for univariate cox regression analysis, lasso regression analysis and multivariate cox regression analysis to identify prognostic-related key genes. This ensured that an improved understanding of the mechanism and prognosis of lung squamous cell carcinoma was obtained.

## 2. Material and methods

### *2.1 Data Collection and data processing*

The Gene Expression Omnibus (GEO, https://www.ncbi.nlm.nih.gov/geo/) database from the National Center for Biotechnology Information was searched for publicly available studies and samples that fulfilled the following criteria for analysis: (1) the gene expression data series contained LUSC tissue and normal tissue samples; (2) the species of the samples was Homo Sapiens. Finally, three gene expression profiles (GSE2088, GSE6044, GSE19188) were collected for further analysis [14]. These three datasets were downloaded, the details of which are shown in Sup.Table 1. GEO2R was used to calculate the fold change value and P-values of every gene in each group. Any gene for which |log2FC| > 1 and P-value < 0.05 was defined as a differentially expressed gene [15]. If multiple probes corresponded to the same gene, the maximum value of fold change was considered the gene expression level.

### *2.2 TCGA data validation*

To ensure accuracy, further validation analysis was performed to confirm the results. LUSC RNAseq data was downloaded from The Cancer Genome Atlas (TCGA, https://www.cancer.gov/tcga) database, acquired using the Illumina Hiseq platform. The dataset included 502 LUSC samples and 49 normal samples. The genomic data used was standardized using the fpkm method [16]. For repetitive genetic data, the largest value was utilized as the expression level. Genes with a mean expression of less than 0.5 were excluded in each case to ensure significantly expressed genes were evaluated. The p-values of differentially-expressed genes were identified using a Wilcox test. From analysis methods of the GEO data, DEGs in the SCC samples compared with normal samples were identified. Raw P-values obtained from the Wilcox test and log2FC calculation were adjusted using the BH16 method [17]. Thresholds of |log2FC| > 1.0 and an adjusted P-value of < 0.05 were selected. The process above was conducted using the R package "limma".

### *2.3 Functional and pathway enrichment analysis*

The GO database has a large collection of gene annotation terms, allowing genome annotation using consistent terminology. GO enrichment analysis including molecular function, cellular components and biological processes, identified which GO terms were over or underrepresented within a given set of genes [18]. The KEGG knowledge database, an integrated database resource, is generally used to identify functional and metabolic pathways [19]. GO and KEGG analysis were both conducted using the Database for Annotation, Visualization and Integrated Discovery (DAVID, https://david.ncifcrf.gov/), analysis tools for extracting meaningful biological information from multiple gene and protein collections [20,21]. Up- and down-regulated genes were analyzed separately, a P-value < 0.05 considered the threshold value.

### *2.4 PPI network construction and analysis of modules*

Protein-protein interaction (PPI) networks can assist in identifying key genes and pivotal gene modules involved in the development of LUSC from an interaction level. The Search Tool for The Retrieval of Interaction Genes (STRING, https://string-db.org/) was used to construct PPI networks. STRING is a database which provides critical assessment and integration of protein–protein interactions, including physical and functional associations [22]. DEGs were mapped to STRING to evaluate the PPI information and set confidence score > 0.4 as the cut-off standard. Cytoscape was used to visualize the PPI network, a practical open-source software tool for the visual exploration of biomolecule interaction networks consisting of protein, gene and other types of interaction [23]. Module analysis was conducted using the plug-in Molecular Complex Detection (MCODE) in Cytoscape to display the biological significance of gene modules [24].

### *2.5 Weighted correlation network analysis of DEGs*

In the WGCNA algorithm [25], the power exponential weighting of gene correlation coefficients was used to represent the correlation between genes. The power value $\beta$ was selected such that connections between genes were subject to scale-free network distribution. A topological matrix that incorporated surrounding genetic information was calculated relative to distance d.

$$S_{ij} = \mathrm{cor}(i, j)$$

$$\alpha_{ij} = |S_{ij}|^{\beta}$$

$$\omega_{ij} = \frac{l_{ij} + \alpha_{ij}}{\min\{k_i, k_j\} + 1 - \alpha_{ij}}$$

$$d_{ij} = 1 - \omega_{ij}$$

where $l_{ij} = \sum u\ \alpha iu\alpha uj$ and $k_i = \sum u\ \alpha iu$ represented node connectivity. The samples were firstly clustered using hierarchical clustering and the threshold set to 40,000 to eliminate outliers. d was then calculated and the dynamic pruning method was used after determining values of gene module parameters (maxBlocksize=7000, deepSplit = 2, minModuleSize = 80, mergeCutHeight = 0.25). After obtaining these, Pearson correlation coefficients of the gene modules and phenotypes (cancer tissue or paracancerous tissue samples) were calculated. This allowed selection of gene modules closely related to tumorigenesis. Genes in the gene modules from GO and KEGG analysis were used to observe the function of each gene module. Next, two indicators for the genes in these gene modules were calculated: one was the Pearson correlation coefficient of the gene in the module and the first principal component of the module, termed Module Membership (MM); the other is the Pearson correlation coefficient of the gene and phenotype in the sample, termed Gene Significance (GS). If a gene within a module had both large MM and GS, then the gene was considered to be a key gene in the module. In this study, genes in which the MM and GS were in the upper quartile of all genes in the module were identified as key genes. The key genes in key modules (modules with a high correlation coefficient for phenotype) were compared with key genes derived from PPI to obtain decisive genes in the development of LUSC. The WGCNA calculations were accomplished using the "WGCNA" package in R [26].

### *2.6 Survival analysis*

Patient clinical data were downloaded from the TCGA website, then samples missing overall survival (OS) data were deleted, finally resulting in a total of 493 samples for survival analysis. Univariate Cox proportional hazards regression analysis was used to screen for genes significantly associated with prognosis ($P < 0.05$) [27]. Later, lasso regression analysis was used to eliminate collinearity between genes. After performing 1000 10-fold cross-validations, the $\lambda$ value in which the error was minimized was selected as the optimum $\lambda$ parameter value [28]. Then multivariate Cox proportional hazards regression analysis was used to find key genes involved in the establishment of a prognostic model [29]. The model used disease risk scores as predictors of prognostic status. The disease risk score was determined by parameter $\beta$ of the multivariate Cox proportional hazards regression analysis and the magnitude of the expression of each gene in the sample, using the following formula:

$$\text{Risk score} = \sum_{i=1}^{N} Expi \times \beta_i$$

where Expi and $\beta$ are the expression levels of gene i in a particular patient and coefficients of the gene i in the multivariate cox regression analysis [30]. The risk score for each patient was calculated, which were then categorized into high or low-risk in comparison with the median value. Kaplan-Meier survival curves were then plotted to evaluate whether the prediction effect of the model was significant ($p<0.05$). The statistical method used in this process was a Log-rank test. The predictive performance of this model at different endpoints (1, 3 or 5 years) was assessed using a time- dependent receiver operating characteristic (ROC) curve [31]. The R packages used in the survival analysis procedure included: "survival", "caret", "glmnet", "survminer" and "survivalROC".

## 3. Results

### *3.1 Identification of DEGs*

Prior to GEO2R analysis, hierarchical clustering was employed to detect sample groups and remove data deviating from the sample group. After measuring the quality of samples, in total there were 97 normal lung samples and 84 with LUSC. Finally, a total of 1013, 675 and 3398 DGEs were identified from the three datasets. Among these DEGs, 128 genes were found in all three sets, including 88 significantly up-regulated DEGs and 4 significantly down-regulated DEGs. In a TCGA dataset containing 49 normal samples and 499 LUSC samples, 3348 up-regulated genes, 3387 down-regulated genes were identified (information on DEGs of the three GEO and one TCGA datasets and the overlap of genes between the datasets is displayed in Supplementary Table). The intersection is shown in Figure 1, including 85 significantly up-regulated and 39 down-regulated genes, the change in direction of expression of TCGA consistent with the DEGs in the GEO database. These genes were used to perform subsequent functional and pathway enrichment analysis.

***3.2 GO term and KEGG pathway enrichment analysis of DEGs***

The function and mechanism of the DEGs identified is this way were further evaluated by GO term and KEGG pathway analysis. Up- and down-regulated genes were uploaded to DAVID to conduct the analysis. In biological processes, up-regulated DEGs were significantly associated with epidermis development, collagen catabolic process, collagen fibril organization. Down-regulated DEGs were predominantly involved in cell adhesion, signal transduction, positive regulation of inflammatory response and regulation of cell growth. For molecular function, significantly up-regulated DEGs were represented by response to protein binding, and structural molecular activity while significantly down-regulated DEGs were involved in flavin adenine dinucleotide binding, thrombospondin receptor activity, and protein binding. For cellular components, significantly up-regulated DEGs were enriched in cytosol, extracellular matrix, desmosome and those that were significantly down-regulated were found in extracellular exosome and space, and membrane raft. Figure 2 presents additional details regarding GO term analysis. In addition, KEGG pathway analysis identified six enriched pathways, in which three were over-represented in up-regulated genes, including cell cycle, ECM-receptor interaction, and amoebiasis. Down-regulated genes enriched in three metabolic pathways, namely drug metabolism of cytochrome P450, tyrosine metabolism and phenylalanine metabolism (Sup.Table 2).

***3.3 PPI network analysis of DEGs***

The PPI network was constructed by Cytoscape based on the STRING database, consisting of 105 nodes and 483 edges, including 76 up- and 29 down-regulated genes (Figure 3A). The 15 DGEs with the greatest degree of connectivity were selected as key genes of SCC. These genes were: RFC4, AURKA, MAD2L1, TPX2, CCNA2, AURKB, TTK, TOP2A, RAD51, CDKN3, TK1, KPNA2, KIF23, PBK and UBE2C, which may play an important role in SCC progression. MCODE in Cytoscape was used to perform module analysis. Four significant modules were identified after MCODE analysis. The most significant module (MCODE score = 23.217) included 24 nodes and 267 edges (Figure 3B). Remarkably, genes in this module were all up-regulated. Functional and pathway enrichment analysis of the DEGs in this module were also conducted using DAVID. GO term enrichment analysis demonstrated that genes in this module were principally enriched in cell division and mitotic nuclear division in biological processes. Cell component analysis indicated that genes were significantly enriched in nucleoplasm, spindle and kinetochore. Molecular functional analysis demonstrated that the genes were principally involved in the binding of ATP and protein. KEGG analysis suggested that the genes were mainly involved in cell cycle (Table 1). The other three modules were shown in Sup. Figure 1.

***3.4 Weighted gene correlation network analysis of DEGs***

Based on the results of hierarchical clustering, we first removed two samples: TCGA.63.5128.01 and TCGA.92.8065.01 (Sup. Figure 2). A value of $\beta=5$ was selected as a soft threshold to establish a gene regulatory network (Figure 4A). After obtaining the gene modules using a dynamic pruning method, it was found that the correlation coefficients of the blue, yellow and turquoise modules were greatest, at 0.538, -0.542 and -0.870, respectively (Figure 4C). In addition, the first principal component of the genes in these modules and the Pearson correlation coefficient between the modules in terms of clustering, were calculated. From these results (Figure 4B), we can see that the three modules turquoise, yellow and red were of greatest consistency. The correlation coefficients of these three modules and phenotypes were negative. The GO and KEGG analysis results for the blue, yellow and turquoise modules are shown in Sup. Figure 3. It can be seen that the genes in the yellow module were more related to inflammatory and immune response processes, also a key module most likely to appear in the WGCNA analysis of other disease and control groups [32]. The blue module was more closely related to cell cycle, mitosis, chromosome structure changes, etc., possibly related to the excessive proliferation of cells during cancer, targeted by many classic anticancer drugs such as paclitaxel and vinblastine [33,34], that play important roles in these processes. Thus, genes in this module are important for drug development. The results of GO and KEGG analysis of the turquoise module were more complex. This principally included cell migration and adhesion, response to hormones, tissue development, cell-growth pathway (TGF-beta signaling pathway, etc.), and immune processes. Among the 20 key genes obtained by PPI, 8 genes were located in the blue module and belonged to key genes of the blue module (Sup.Table 3)., while the other 11 genes were not present among key genes of any module.

### *3.5 Survival analysis*

In order to establish an effective model for predicting prognostic status, univariate Cox proportional hazards regression analysis was used, with Lasso regression analysis and multivariate Cox proportional hazards regression analysis to screen genes. In the Univariate Cox proportional hazards regression analysis, 111 genes with significant effects on prognosis ($p<0.01$, Supplementary Table) were identified. In lasso regression, following 10-fold cross-validation with 1000 repeats, $\lambda$ was 0.0007079 (Figure 5D), where Partial Likelihood

Deviance was smallest. For a λ of this value, 33 of the 111 genes had coefficients that were not zero (Figure 5E). A total of 20 genes were obtained by the forward-backward selection technique in multivariate Cox proportional hazards regression analysis for to establish a prognostic risk score model, namely: KLK8, POPDC3, IGLL1, PLEKHA6, PTGIS, ANKFN1, TM4SF19, JAG1, OR2W3, RALGAPA2, RPH3AL, PCDHGA7, C10orf55, ACPT, MMP12, H1FOO, CT45A10, UGT2A1, a low expression of which improved prognosis, and LYSMD1 and ONECUT3, where improved prognosis occurred with high expression (Figure 6A).

Risk score = (0.00943 × KLK8) + (0.01536 × POPDC3) + (0.04825 × IGLL1) + (−0.04272 × LYSMD1) + (0.01822 × PLEKHA6) + (0.01318 × PTGIS) + (0.14178 × ANKFN1) + (0.02925 × TM4SF19) + (0.00116 × JAG1) + (0.11625 × OR2W3) + (−0.47624 × ONECUT3) + (0.01810 × RALGAPA2) + (0.03305 × RPH3AL) + (0.06967 × PCDHGA7) + (0.06217 × C10orf55) + (0.08721 × ACPT) + (0.00082 × MMP12) + (0.58094 × H1FOO) +(0.01348 × CT45A10) + (0.02889 × UGT2A1).

The distribution of risk scores and groupings are shown in Figure 5A. Patient risk scores are ranked and patient gene expression plotted as a heatmap (Figure 5C). It can be seen from the heat map that gene expression levels with a negative coefficient gradually decreased as risk score increased. Conversely, expression levels of genes with a positive coefficient displayed the opposite. As risk scores increased, the number of patients that died increased and duration of survival gradually decreased, confirming that the risk model was realistic (Figure 5B). The Kaplan–Meier curves were grouped by defined risk scores. It can be seen that prognosis of the low-risk group was significantly better than that of the high-risk group (Figure 5F). By predicting survival of patients at 1, 3, and 5 years, the area under the curve (AUC) of the ROC curves obtained from the risk-based prediction model was 0.828, 0.826 and 0.824 (Figure 6B).

**4. Discussion**

Although many relevant studies of LUSC have been performed, early diagnosis, efficacy of treatment and prognosis for LUSC remain poorly resolved. For diagnosis and treatment, it is necessary to further understand the molecular mechanisms resulting in occurrence and development. Due to the development of high-throughput sequencing technology, the genetic alterations due to disease progression can be detected, indicating gene targets for diagnosis, therapy and prognosis of specific diseases.

In this study, the intergrading of GEO and TCGA data for LUSC resulted in the identification of 124 significantly DEGs in lung squamous cell cancer compared with normal samples, including 85 up-regulated DEGs and 39 that were down-regulated. GO analysis demonstrated that the up-regulated genes were principally involved in epidermal development and DNA unwinding in DNA replication, while the down-regulated genes were mainly enriched in cell adhesion, signal transduction and positive regulation of inflammatory response. KEGG analysis indicated that these differentially expressed genes were involved in cell cycle regulation, cancer-related pathways, ECM receptor interaction and various metabolic pathways. In addition, 15 genes of high significance, including RFC4, AURKA, MAD2L1, TPX2, CCNA2, AURKB, TTK, TOP2A, RAD51, CDKN3, TK1, KPNA2, KIF23, PBK and UBE2C, were identified by constructing a PPI network. Through module analysis we identified four significant gene modules, in which the most significant module contained 26 nodes and 267 edges strongly associated with cell division and related activities. Interestingly, the genes in this module were all up-regulated and fifteen key genes of the highest significance were contained in this module. Through WGCNA analysis, three significant gene modules, in which the blue module was positively correlated with the development of tumors, the opposite of the others, were obtained. Key genes identified in the PPI network were mostly contained in the blue module, including AURKA, RAD51, TTK, AURKB, CCNA2, TPX2, KPNA2 and KIF23. GO analysis indicated that the blue module was associated with cell cycle, consistent with PPI submodule analysis. The WGCNA package constructed a network based on the correlation between genes, whereas the PPI network was based on protein networks reported in the known literature. It seems appropriate to combine WGCNA and PPI methods to identify key genes. Thus, these eight genes were selected as hub genes for the occurrence and development of LUSC. All eight are involved in cell cycle progression and regulate different mitotic events. A large number of studies have reported that abnormal expression levels were found in multiple types of human malignancy and have potential as anticancer therapeutic targets [35-40].

AURKA and AURKB are members of the aurora kinase family and both play central roles in regulating cell-cycle progression from G2 through to cytokinesis. AURKA is involved in many mitotic events, including centrosome maturation, mitosis entry, mitotic spindle formation and cytokinesis [41]. AURKB exerts its function by regulating chromosomal alignment, segregation and cytokinesis, as the catalytic protein of the chromosomal passenger complex (CPC) [40]. Yu *et al.* identified functional interaction between AURKA and AURKB, assisting in protection of their stability and partially explaining their persistent high expression and activity in cancers [43]. Ma *et al.* verified that up-regulation of miR-32 down-regulated AURKA and thus inhibited the occurrence and development of NSCLC [44]. Jin *et al.* performed a bioinformatics analysis and proposed that AURKB may be the key gene in LUAC and could result in poor prognosis [45].

RAD51 is the central protein in the homologous recombination (HR) pathway and is critical for DNA replication and repair. Recent studies have reported that RAD51 is also implicated in the suppression of innate immunity [46]. Hu *et al.* pointed out that RAD51 expression was elevated by a KRAS mutation which predominantly occurred in lung adenocarcinomas. Furthermore, they found that the high expression of RAD51 was associated with poor survival in lung adenocarcinoma through analysis of data from the TCGA database [47]. Moreover, expression levels of RAD51 were also linked to resistance to chemotherapy. Takenaka *et al.* reported that the combined expression of RAD51 and another DNA repair enzyme, ERCC1 (Excision repair cross-complementation group 1) is associated with resistance to platinum agents, indicating the potential role of RAD51 in prognosis [48].

Monopolar spindle1 (Mps1), also known as TTK, is a dual-specificity kinase that phosphorylates serine/threonine and tyrosine residues. TTK is a key mitotic checkpoint protein, similar to AURKA and AURKB. It functions as a critical component of the SAC (spindle assembly checkpoint) signaling cascade, ensuring correct chromosome segregation [49]. High expression of TTK in LUAC and LUSC has been revealed through analysis of TCGA data, where it eliminates lung cancer by augmenting apoptosis and polyploidy [50]. Cyclin family member CCNA2 is involved in cytoskeletal dynamics, epithelial–mesenchymal transition (EMT) and metastasis [51]. Chen *et al.* used two lung adenocarcinoma cell lines to verify that miR-137 can induce G1/S cell cycle arrest and dysregulate mRNA expression in cell cycle associated proteins, including CCNA2 [52].

TPX2, a microtubule-associated protein, is required for the formation of normal bipolar spindles and chromosome separation. Overexpression of TPX2 can induce abnormal centrosome amplification, leading to aneuploidy formation and malignant transformation of cells. It can also promote cell proliferation and participate in cell cycle and apoptosis regulation [53]. It is overexpressed in LUSC, which has been specifically verified [54]. Remarkably, TPX2 can specifically interact with AURKA to complete a series of biological functions including targeting of AURKA to the spindle microtubes and assembly of spindles of the correct length which faithfully segregate chromosomes [55]. Co-expression of TPX2 and AURKA have been verified, depletion of either causing defects in the formation of spindles [35,55]. In LUSC, up-regulation of AURKA and TPX2 are correlated, a high correlation in tumor tissues having been established [56].

KPNA2 regulates the transportation of a variety of important proteins from the cytoplasm into the nucleus. Increasing evidence has shown that KPNA2 contributes to the process of gene regulation, including that of oncogenes and tumor-suppressor genes, further affecting biological functions, such as cell proliferation, differentiation, cell-matrix adhesion, colony formation and migration [57]. Wang *et al.* found that KPNA2 was overexpressed in the nuclei of tumor cells in NSCLC, through analysis of the cancer cell secretome and tissue transcriptome, suggesting that KPNA2 is a potential biomarker for NSCLC [58].

Nuclear protein KIF23 is reported to interact with the aurora kinase family to regulate cell division and cytokinesis [59]. A reduction in KIF23 gene expression was able to suppress cell proliferation [60]. A number of experimental and bioinformatics studies have pointed to high levels of KIF23 expression observed in the majority of metastatic lung cancer tissues, indicating that KIF23 plays a role in NSCLC occurrence and development [36,61-63]. Importantly, abnormal behavior of genes in lung cancer has been widely reported. However, its role and molecular mechanisms in LUSC require additional investigation.

We have established a prognostic model that predicts patient survival. This model contains 20 key genes, some of which have not been reported in the literature in LUSC studies, including POPDC3, IGLL1, TM4SF19, RALGAPA2, RPH3AL, PCDHGA7 or ACPT. The present study found that KLK8 (kallikrein8) is involved in the development of a variety of tumors, including cervical, ovarian and breast cancer. KLK8 is expressed in different tissues and cells and regulated by steroid hormones [64-66]. KLK is distributed principally in normal tissues such as the kidneys, breast, skin, tonsil, testis, *etc.*, but KLK8 is expressed less in the lungs, which may be a reason for the small coefficient (0.00943) for KLK8 in the risk score [67]. That study demonstrated that KLK8 is highly expressed in ovarian cancer and can be used as a marker for tumorigenesis. Whether KLK8 can be used as a biomarker for squamous cell carcinoma deserves further investigation [68]. High expression of LYSMD1 and ONECUT3 has a protective effect on squamous cell carcinoma. LYSMD1 is a member of the LysM domain. The LysM domain is widely found in plants, bacteria, fungi and animals [69,70] and can be combined with a variety of other proteins. At present, few studies have reported on its role in mammalian tissues, but it can be seen from the results of survival analysis that high expression of LYSMD1 may have a protective effect on keratinized squamous cell carcinoma.

The ONECUT transcription factor contains three members in mammalian cells, namely hepatocyte nuclear factor 6 (ONECUT-1), ONECUT-2 (OC-2) and ONECUT-3 (OC-3), roles of which overlap [71]. ONECUT participates in the biological functions of cell cycle and proliferation regulation, cell differentiation, organ formation, cell migration, cell adhesion and metabolism by regulation of gene expression [72]. Current studies on the regulation of cell proliferation by OC transcription factors are focusing mainly on ONECUT-1. It is worth noting that ONECUT-1 has a certain tissue specificity for the regulation of cell proliferation. For example, overexpression of ONECUT-1 in colonic adenocarcinoma cell line Caco-2 inhibits cell cycle progression [73]. At present, there are few reports of the regulation of cell cycle in ONUCUT-3, but the high expression of this gene has a significant prognostic effect. In lung tissue, the expression of ONECUT-1 is reduced, so ONECUT-3 may regulate cell cycle and adhesion, but this requires further experimental verification. Related studies have shown that the occurrence of lung cancer and the metastasis of cancer cells can significantly change the patient's olfactory function, consistent with the genes we have identified. Thus,

OR2W3 (olfactory receptor family 2 subfamily W member 3) could potentially be used as a tumor marker for the examination of squamous cell carcinoma [74,75].

H1FOO is an H1.8 linker histone. Expression levels of this gene are closely related to time course, but different from other H1 protein subtypes. It not only functions in oocyte-embryo development transition, but also plays a crucial role in the genome reprogramming process. When the expression of the H1FOO gene in oocytes increases, the pluripotency of the cells also increases, indicating that H1FOO is involved in chromosome remodeling and gene reprogramming. This process is closely related to the occurrence of cancer, so to some extent, the levels of H1FOO expression reflect the degree of cancer cell proliferation [76,77].

Ankyrin repeats play an important role in the progression of inflammation-associated tumors. At present, it has been found that ankyrin repeats are highly expressed in cervical cancer, liver cancer, cholangiocarcinoma, breast cancer, *etc.* and that their expression levels are closely related to depth of tumor invasion, grading stage [78] and tumor metastasis. Studies have suggested that ankyrin repeats may play a role in cell proliferation by regulating the function of the important tumor suppressor genes p53, Rb and hypoxia inducible factor-1 (HIF-1), and the IL-8 pathway [79]. The protein fibronectin is a glycoprotein secreted by tumor-associated fibroblasts and tumor cells. Deposition of fibronectin stimulates cancer cells to bind to insulin-like growth factor binding protein-3 (IGFBP-3) and epidermal growth factor (EGF) [80]. When fibronectin increases in concentration, IGFBP-3 and EGF promote cancer cell proliferation. Conversely, if fibronectin is deleted, cancer cell proliferation is inhibited. This is consistent with the large positive coefficient (0.14178) for the ANKFN1 (ankyrin repeat and fibronectin type III domain containing 1) gene obtained in the present study.

## 5. Conclusions

In the present study, microarray data from the GEO database were integrated with RNA sequencing data from the TCGA, to identify key genes and more important hub genes. Finally, we identified eight hub genes associated with the pathogenesis and progression of LUSC. Additionally, we performed survival analysis and built a Cox proportional hazards model to identify prognostic biomarkers. A prognostic gene signature consisting of 20 genes was constructed with good performance in predicting overall survival. These results will serve as a reference for future research on the pathogenesis and drug treatment for LUSC. Nevertheless, the lack of experimental verification is a limitation of this study. The predictions obtained from the bioinformatics analysis can be verified by future experimental studies.

**Data availability**

Gene Expression Omnibus (GEO, https://www.ncbi.nlm.nih.gov/geo/)

The Cancer Genome Atlas (TCGA, https://www.cancer.gov/tcga)

Database for Annotation, Visualization and Integrated Discovery (DAVID, https://david.ncifcrf.gov/),

The Search Tool for The Retrieval of Interaction Genes (STRING, https://string-db.org/)

**Abbreviations**

| | |
|---|---|
| SCLC | Small Cell Lung Cancer |
| NSCLC | Non-Small Cell Lung Cancer |
| LUAD | Lung Adenocarcinoma |
| LUSC | Lung Squamous Cell Carcinoma |
| DEGs | Differentially Expressed Genes |
| PPI | Protein-protein interaction |
| FC | Fold Change |
| GEO | Gene Expression Omnibus |
| TCGA | The Cancer Genome Atlas |
| GO | Gene Ontology |
| KEGG | Kyoto Encyclopedia of Genes and Genomes |
| DAVID | Database for Annotation, Visualization and Integrated Discovery |
| STRING | Search Tool for The Retrieval of Interaction Genes |
| MCODE | Molecular Complex Detection |
| WGCNA | Weighted Gene Co-Expression Network Analysis |
| GS | Gene Significance |
| MM | Module Membership |
| OS | Overall Survival |

**Gene Abbreviation**

| | |
|---|---|
| RFC4 | Replication Factor C Subunit 4 |
| MAD2L1 | Mitotic Arrest Deficient 2 Like 1 |
| AURKA | Aurora Kinase A |
| CDKN3 | Cyclin Dependent Kinase Inhibitor 3 |
| RAD51 | DNA Repair Protein RAD51 Homolog 1 |
| TOP2A | DNA Topoisomerase II Alpha |
| TTK | Threonine and Tyrosine Kinase |
| AURKB | Aurora Kinase B |
| CCNA2 | Cyclin A2 |
| TPX2 | Xenopus Kinesin-Like Protein 2 |
| TK1 | Thymidine Kinase 1 |
| PBK | PDZ Binding Kinase |
| KPNA2 | Karyopherinα2 |
| KIF23 | Kinesin Family Member 23 |
| KLK8 | Kallikrein8 |
| OR2W3 | Olfactory Receptor Family 2 Subfamily W Member 3 |
| H1FOO | H1.8 Linker Histone |
| ANKFN1 | Ankyrin Repeat and Fibronectin Type III Domain Containing 1 |

**Funding**

This work was supported by National key research and development program of China (2018YFA0900200), National Natural Science Foundation of China Grants (31771519, 31871083) and Beijing Natural Science Foundation (5182012, 7182087), the Ministry of Science and Technology of China Grant 2015CB559200.Funding for open access charge: National key research and development program of China.

**Author Contributions Statement**

All authors contributed to the work presented in this paper. Miaomiao Gao and Zhengwei Xie identified the research topic. Miaomiao Gao performed the selection of GEO dataset data, the screening of GEO differential expressed genes, the establishment of PPI, the GO and KEGG analysis of DEGs, drawing of differentially expressed genes and wrote relevant content. Weilaixin Kong performed the screening of TCGA differential expressed genes, WGCNA analysis, survival analysis, GO and KEGG analysis of gene modules and wrote relevant content. Zhuo Huang and Zhengwei Xie gave suggestions on the choice of experimental methods and the writing of the paper. The manuscript was reviewed and approved by all authors.

## Conflicts of interest

None.

**Figure captions**

**Figure 1.** Overview of microarray signatures. (A-D). Volcano plots of DEGs of GSE2088, GSE6044, GSE19188 and TCGA datasets. The red and blue points in the plot represent statistically significant up and down-regulated DEGs. (E). Venn diagrams of DEGs of the three GEO and one TCGA dataset.

**Figure 2.** GO enrichment analysis of DEGs. (A) Up-regulated genes. (B) Down-regulated genes. The y-axis shows significantly enriched GO terms, and x-axis represents significance of enrichment.

**Figure 3.** Protein-protein interaction network analysis of DEGs. (A) PPI network of 124 significantly differentially expressed genes. Red: up-regulated genes; Green: significantly down-regulated genes; (B) The most significant module of the PPI network. Node size is positively related to degree of expression.

**Figure 4.** WGCNA analysis of DEGs within the TCGA data. (A). Left panel: Abscissa: power value; ordinate: R2 value after linear regression of -log10 (k) and log10 (P (k)). k: connectivity of the gene nodes; P (k): probability of such a node. Horizontal line: 0.9. Right panel: Abscissa: power value; ordinate: mean connectivity of gene nodes. (B). Interrelationship between different gene modules. Colors in the heat map represent Pearson correlation coefficients between gene modules. Expression levels of gene modules are represented by the first principal component. (C). Gene modules and phenotypes quantified using Boolean variables (1 represents occurrence, 0 represents no occurrence) were used to calculate correlation coefficients, represented as a heat map. p- values are displayed in brackets. (D). Blue, yellow and turquoise module gene correlation scatter plots. X-axis represents molecule membership, i.e. Pearson correlation coefficients of gene and module (MM). Y-axis represents the importance of the gene for the phenotype, i.e. Pearson's correlation coefficient of gene and phenotype (GS: phenotype is represented by a Boolean

variable). In the yellow, blue, turquoise modules, the upper quartile value of MM was 0.756, 0.683 and 0.708 respectively; the upper quartile value of GS was 0.431, 0.390 and 0.631, respectively.

**Figure 5.** Lasso regression and multivariate cox regression related results. (A). Distribution of risk scores. The abscissa is arranged in order of patient risk score, with y-axis representing risk score, with more than 10 records plotted as 10. Patient with a score less than the median is plotted in green, those equal to or greater than the median in red. (B) Distribution of duration of survival. The abscissa is arranged in order of patient risk score and ordinate representing patient survival time. (C) Heat map of gene expression levels of the most 20 significant genes. Patient risk scores increased from left to right. (D). Results of Lasso regression 1000 times 10-fold cross validation. λ was determined when partial likelihood deviance was smallest. (E). Coefficient curve. Different colored lines represent coefficient sizes of individual genes in different cases. The abscissa represents log (λ) and the number of coefficients (top) that are not zero under this penalty factor. (F). Survival curve for patients with different risk scores. Patients were divided into two groups according to the median survival curve score. Blue represents patients with a lower risk score. p- value $<0.001$.

**Figure 6.** (A). Forest plot for multivariate cox regression. The HR value is eβ. This figure displays the 95% confidence interval for the HR value over the box plot with associated p-values. An HR of less than 1 indicates that high gene expression improves prognosis, and HR greater than 1 the opposite. (B). ROC curves for the model representing 1, 3 and 5-year predictions, from left to right, respectively. The values in brackets are the areas under the curve.

**Figures**

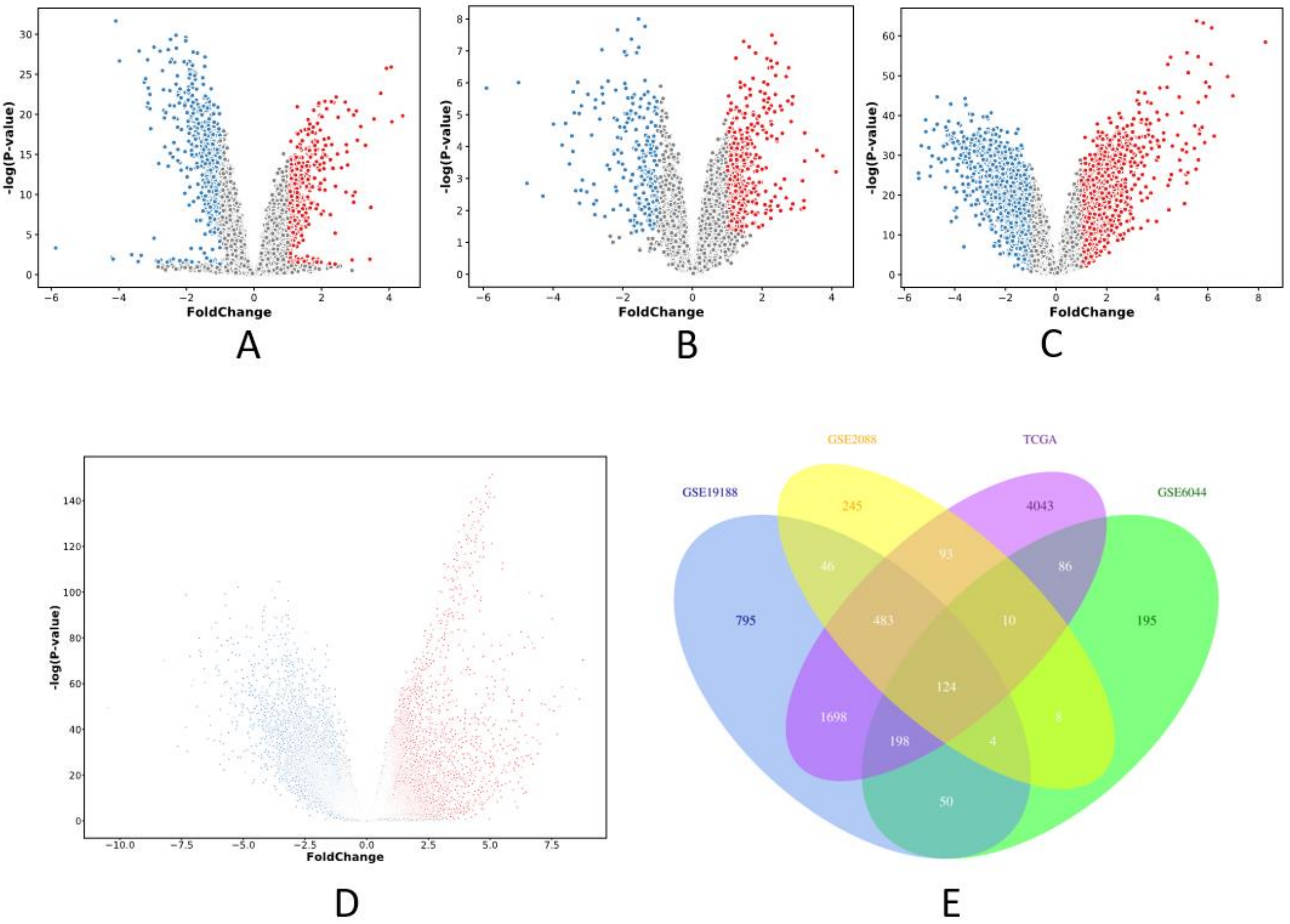

**Figure 1**

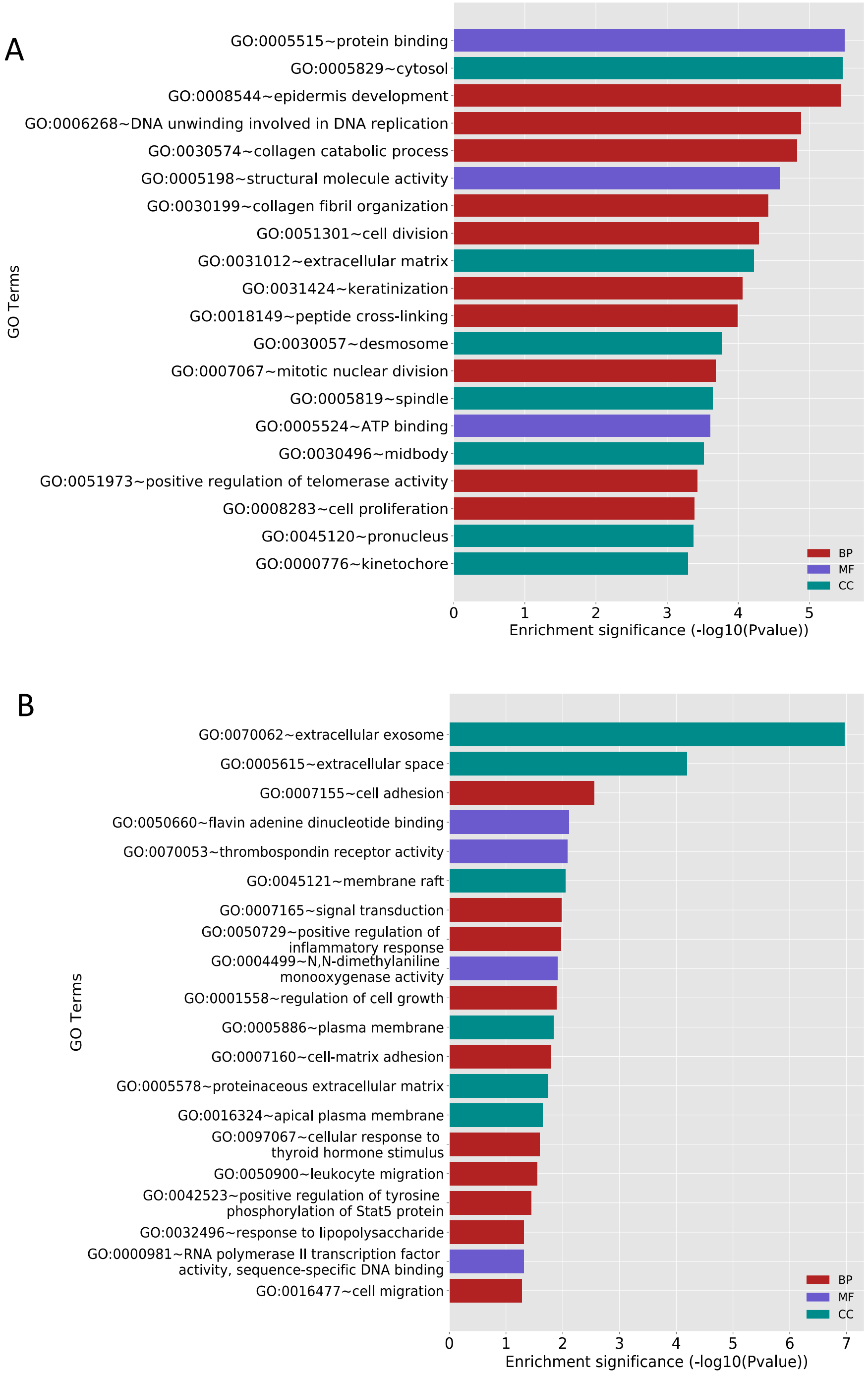
A
GO:0005515~protein binding
GO:0005829~cytosol
GO:0008544~epidermis development
GO:0006268~DNA unwinding involved in DNA replication
GO:0030574~collagen catabolic process
GO:0005198~structural molecule activity
GO:0030199~collagen fibril organization
GO:0051301~cell division
GO:0031012~extracellular matrix
GO:0031424~keratinization
GO:0018149~peptide cross-linking
GO:0030057~desmosome
GO:0007067~mitotic nuclear division
GO:0005819~spindle
GO:0005524~ATP binding
GO:0030496~midbody
GO:0051973~positive regulation of telomerase activity
GO:0008283~cell proliferation
GO:0045120~pronucleus
GO:0000776~kinetochore
GO Terms
BP
MF
CC
0
1
2
3
4
5
Enrichment significance (-log10(Pvalue))
B
GO:0070062~extracellular exosome
GO:0005615~extracellular space
GO:0007155~cell adhesion
GO:0050660~flavin adenine dinucleotide binding
GO:0070053~thrombospondin receptor activity
GO:0045121~membrane raft
GO:0007165~signal transduction
GO:0050729~positive regulation of inflammatory response
GO:0004499~N,N-dimethylaniline monooxygenase activity
GO:0001558~regulation of cell growth
GO:0005886~plasma membrane
GO:0007160~cell-matrix adhesion
GO:0005578~proteinaceous extracellular matrix
GO:0016324~apical plasma membrane
GO:0097067~cellular response to thyroid hormone stimulus
GO:0050900~leukocyte migration
GO:0042523~positive regulation of tyrosine phosphorylation of Stat5 protein
GO:0032496~response to lipopolysaccharide
GO:0000981~RNA polymerase II transcription factor activity, sequence-specific DNA binding
GO:0016477~cell migration
GO Terms
BP
MF
CC
0
1
2
3
4
5
6
7
Enrichment significance (-log10(Pvalue))

**Figure 2**

A

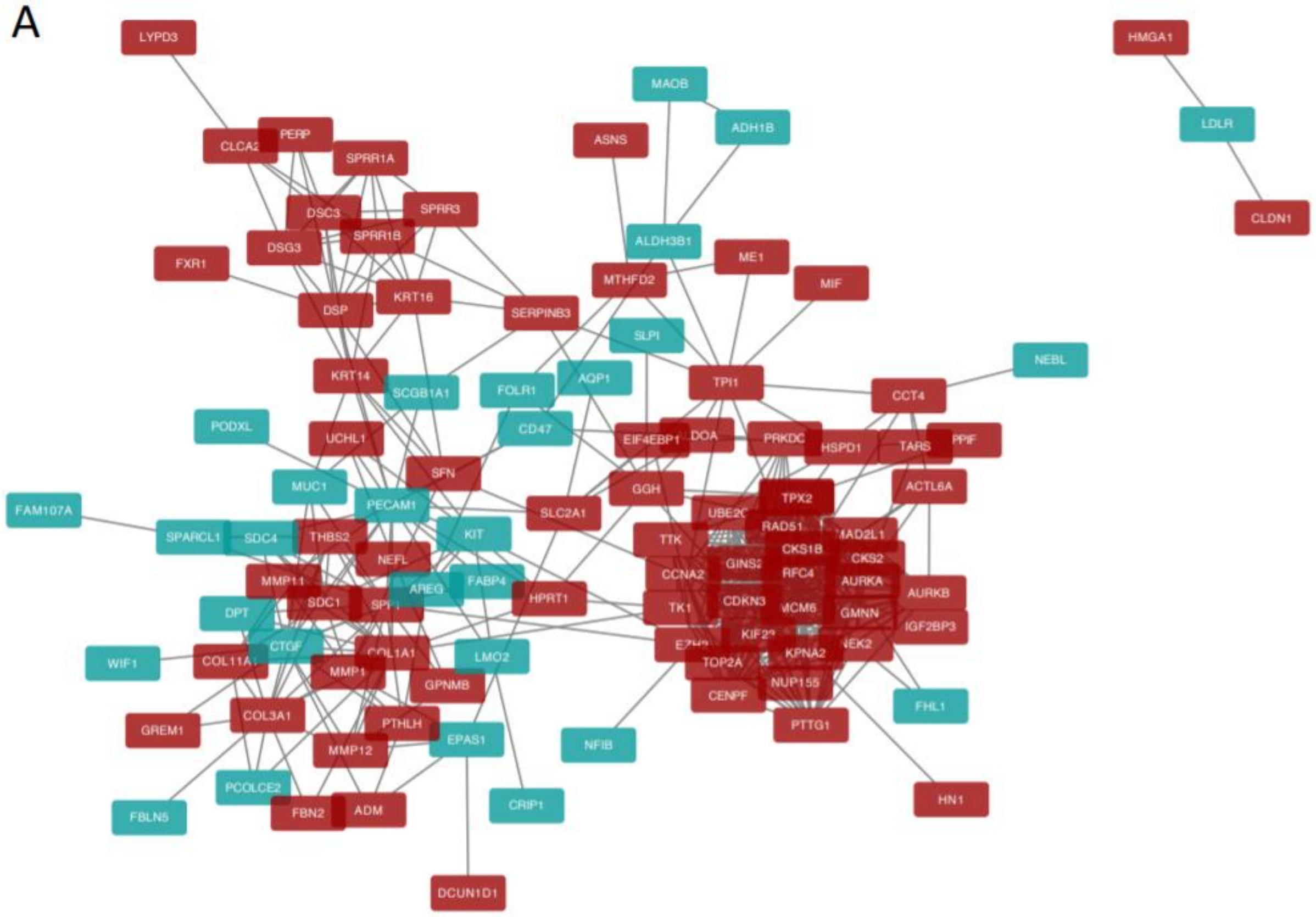

B

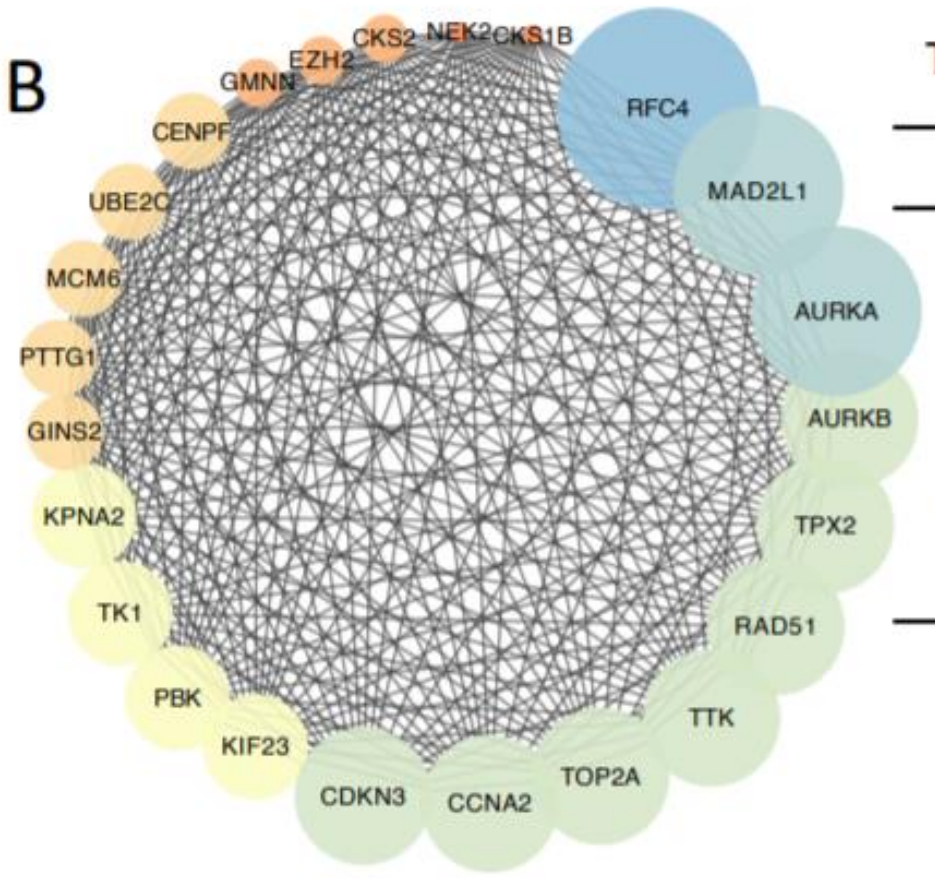

Table 1. GO term enrichment analysis of DEGs in modules.

| GO Term | Count | % | PValue | Genes |
|---|---|---|---|---|
| GO:0051301~cell division | 10 | 41.6667 | 4.23E-10 | CKS1B, MAD2L1, NEK2, CKS2, TPX2, CENPF, AURKA, PTTG1, UBE2C, CCNA2 |
| GO:0007067~mitotic nuclear division | 8 | 33.3333 | 2.82E-08 | NEK2, TPX2, CENPF, AURKA, PBK, PTTG1, AURKB, CCNA2 |
| GO:0031145~anaphase-promoting complex-dependent catabolic process | 5 | 20.8333 | 3.75E-06 | MAD2L1, AURKA, PTTG1, AURKB, UBE2C |
| GO:0042787~protein ubiquitination involved in ubiquitin-dependent protein catabolic process | 5 | 20.8333 | 5.13E-05 | MAD2L1, AURKA, PTTG1, AURKB, UBE2C |
| GO:0006268~DNA unwinding involved in DNA replication | 3 | 12.5 | 8.02E-05 | TOP2A, RAD51, MCM6 |

**Figure 3**

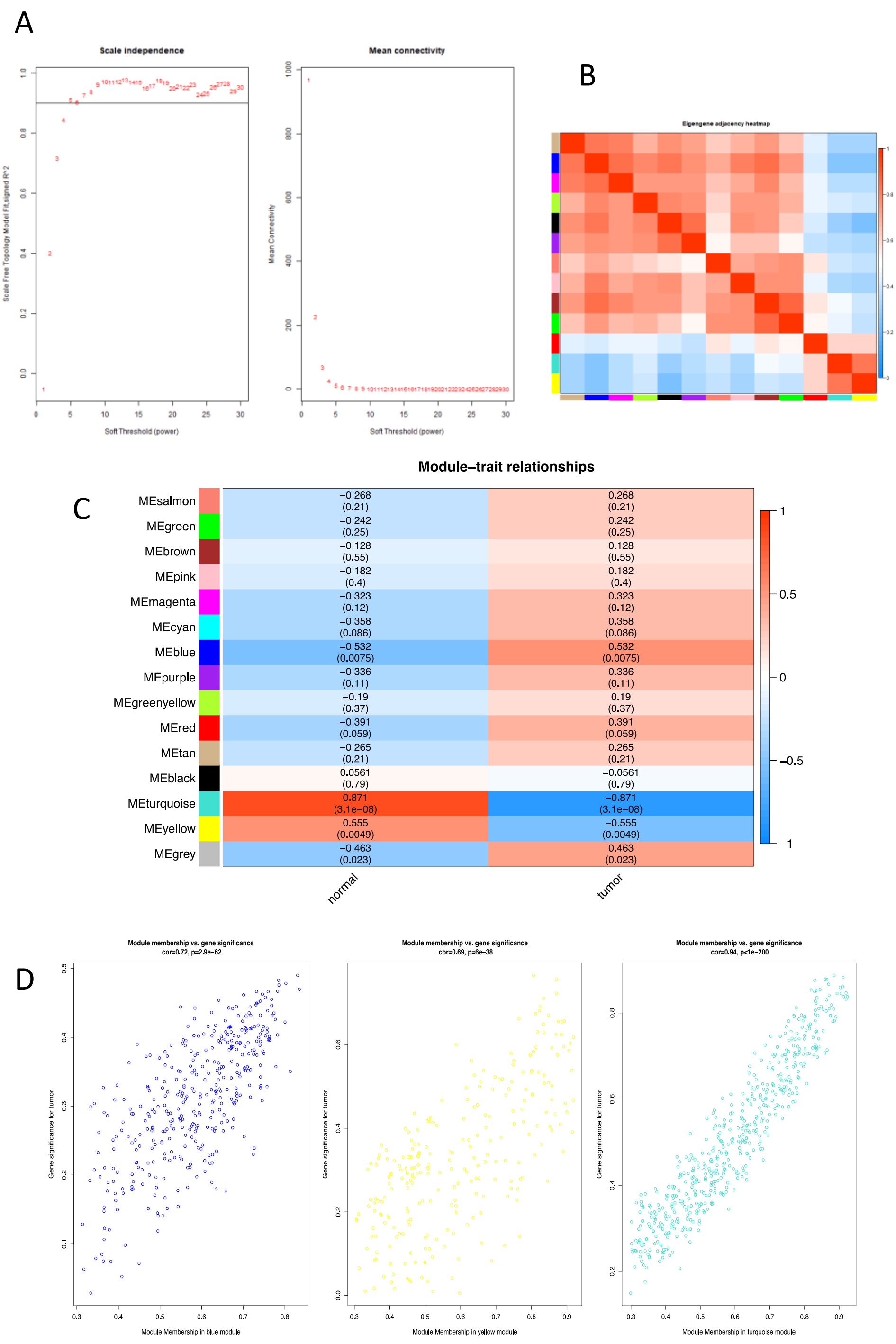
A
Scale independence
Scale Free Topology Model Fit,signed R^2
Soft Threshold (power)
Mean connectivity
Mean Connectivity
Soft Threshold (power)
B
Eigengene adjacency heatmap
C
Module−trait relationships
MEsalmon −0.268 (0.21) 0.268 (0.21)
MEgreen −0.242 (0.25) 0.242 (0.25)
MEbrown −0.128 (0.55) 0.128 (0.55)
MEpink −0.182 (0.4) 0.182 (0.4)
MEmagenta −0.323 (0.12) 0.323 (0.12)
MEcyan −0.358 (0.086) 0.358 (0.086)
MEblue −0.532 (0.0075) 0.532 (0.0075)
MEpurple −0.336 (0.11) 0.336 (0.11)
MEgreenyellow −0.19 (0.37) 0.19 (0.37)
MEred −0.391 (0.059) 0.391 (0.059)
MEtan −0.265 (0.21) 0.265 (0.21)
MEblack 0.0561 (0.79) −0.0561 (0.79)
MEturquoise 0.871 (3.1e−08) −0.871 (3.1e−08)
MEyellow 0.555 (0.0049) −0.555 (0.0049)
MEgrey −0.463 (0.023) 0.463 (0.023)
normal
tumor
D
Module membership vs. gene significance
cor=0.72, p=2.9e−52
Module Membership in blue module
Module membership vs. gene significance
cor=0.69, p=6e−38
Module Membership in yellow module
Module membership vs. gene significance
cor=0.94, p<1e−200
Module Membership in turquoise module
Gene significance for tumor

**Figure 4**

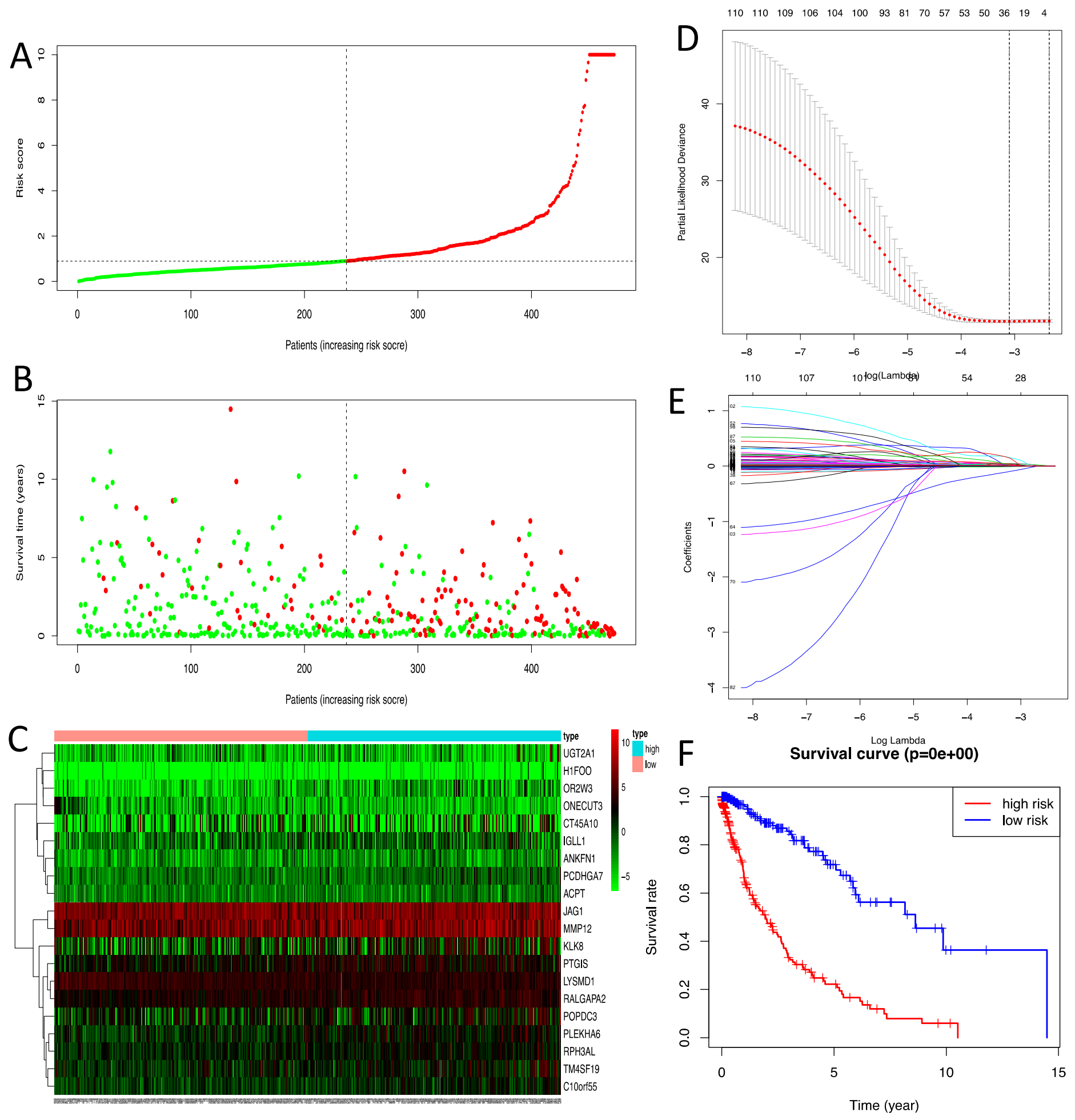

# Figure 5

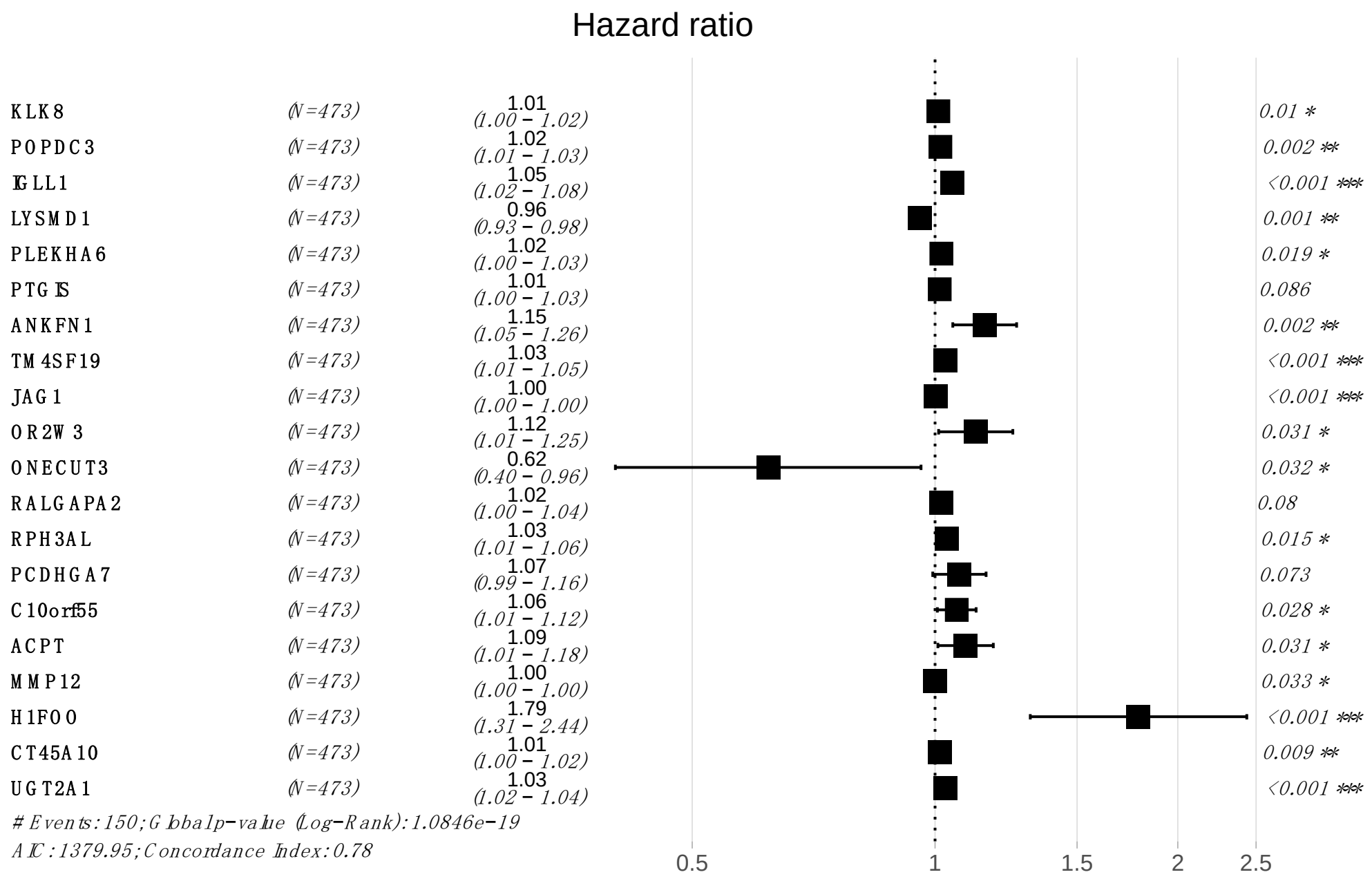

# Figure 6

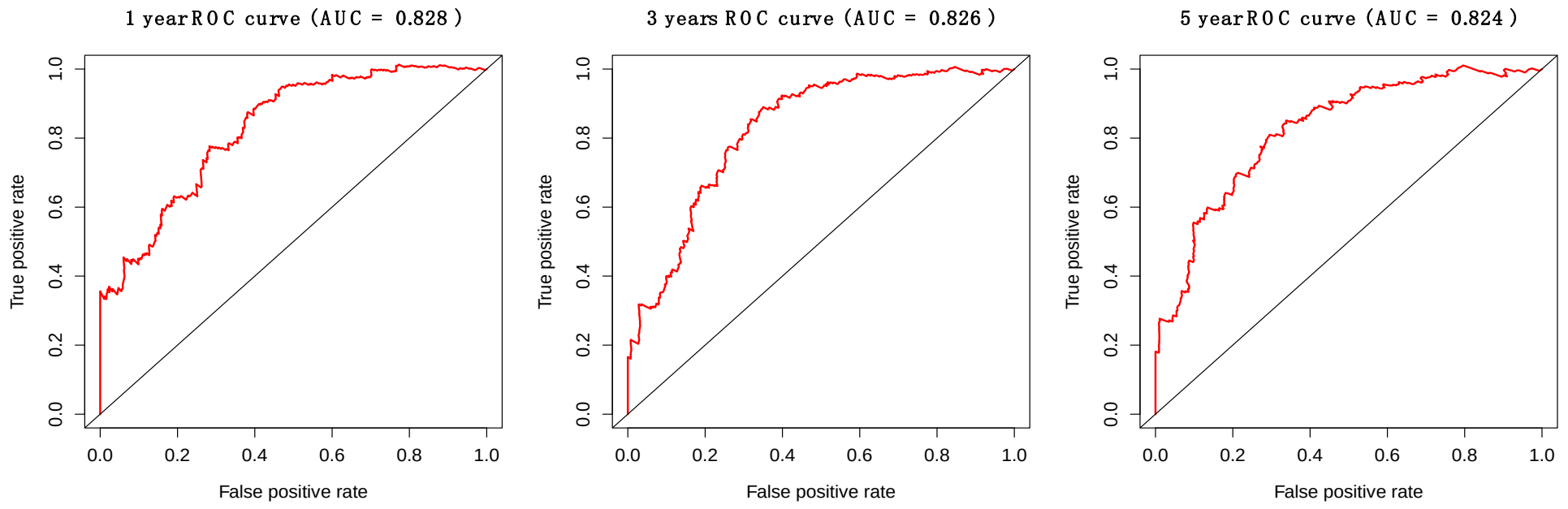

# Figure 7

# Supplementary materials

**Table 1.** Detailed information of the GEO datasets.

| GEO Accession | Platform | Number of LUSC sample | Number of normal sample | Total |
|---|---|---|---|---|
| GSE2088 | GPL962 | 48 | 28 | 76 |
| GSE6044 | GPL201 | 9 | 4 | 13 |
| GSE19188 | GPL570 | 27 | 65 | 92 |
| In all | - | 84 | 97 | 181 |

**Table 2.** KEGG pathway enrichment analysis of DEGs.

| Term | Count | % | P-value | Genes |
|---|---|---|---|---|
| **Up-regulated** | | | | |
| hsa04110: Cell cycle | 7 | 8.2353 | 0.0002 | MAD2L1, TTK, PRKDC, PTTG1, SFN, CCNA2, MCM6 |
| hsa04512:ECM-receptor interaction | 6 | 7.0588 | 0.0004 | SDC1, COL3A1, COL1A1, COL11A1, THBS2, SPP1 |
| hsa05146: Amoebiasis | 4 | 4.7059 | 0.0413 | COL3A1, COL1A1, SERPINB3, COL11A1 |
| **Down-regulated** | | | | |
| hsa00982:Drug metabolism-cytochrome P450 | 5 | 0.0872 | 3.03E-05 | FMO5, FMO3, MAOB, ADH1B, ALDH3B1 |
| hsa00350:Tyrosine metabolism | 3 | 0.0523 | 0.0041 | MAOB, ADH1B, ALDH3B1 |
| hsa00360: Phenylalanine metabolism | 2 | 0.0349 | 0.0460 | MAOB, ALDH3B1 |

**Table 3.** MM and GS scores of eight significant genes.

| | AURKA | CCNA2 | TPX2 | TTK | AURKB | RAD51 | KIF23 | KPNA2 |
|---|---|---|---|---|---|---|---|---|
| MM | 0.732 | 0.699 | 0.69 | 0.738 | 0.7 | 0.74 | 0.8 | 0.776 |
| GS | 0.469 | 0.445 | 0.398 | 0.412 | 0.43 | 0.475 | 0.5 | 0.461 |

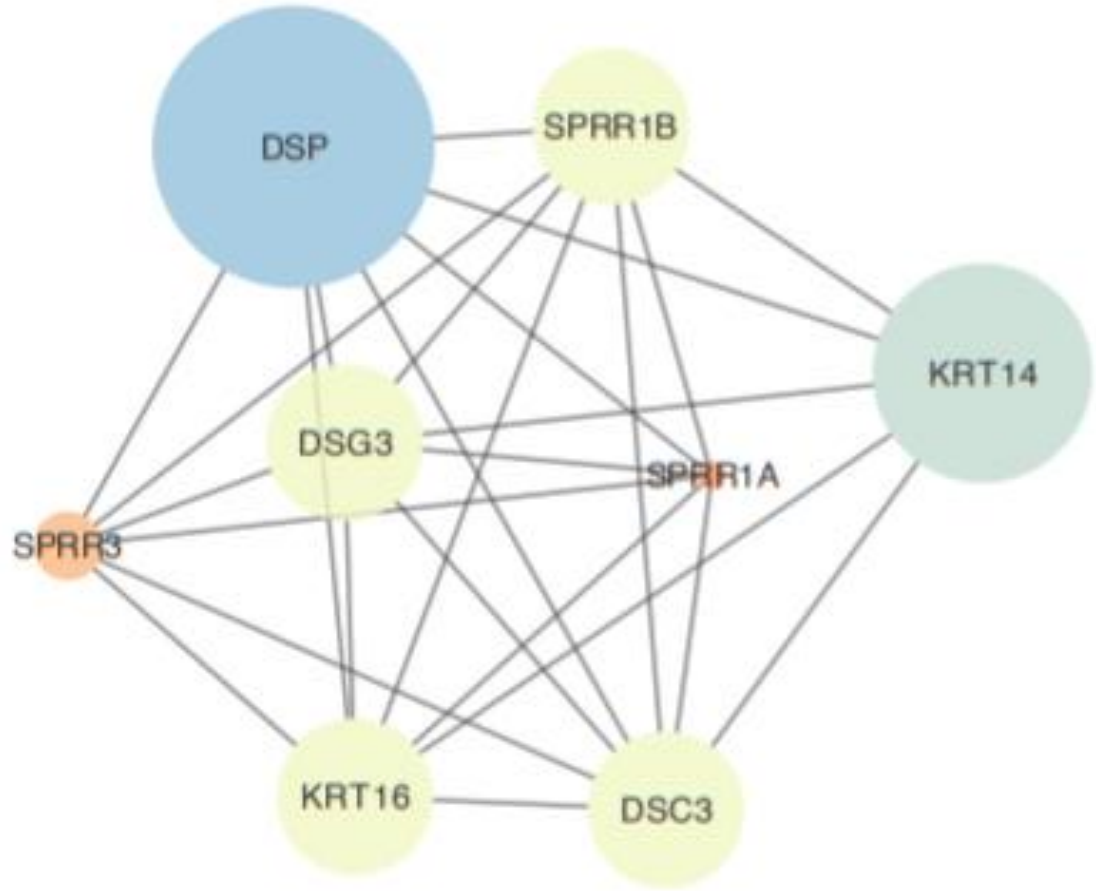

Cluster 2

MCODE score = 7.429

8 nodes 26 edges

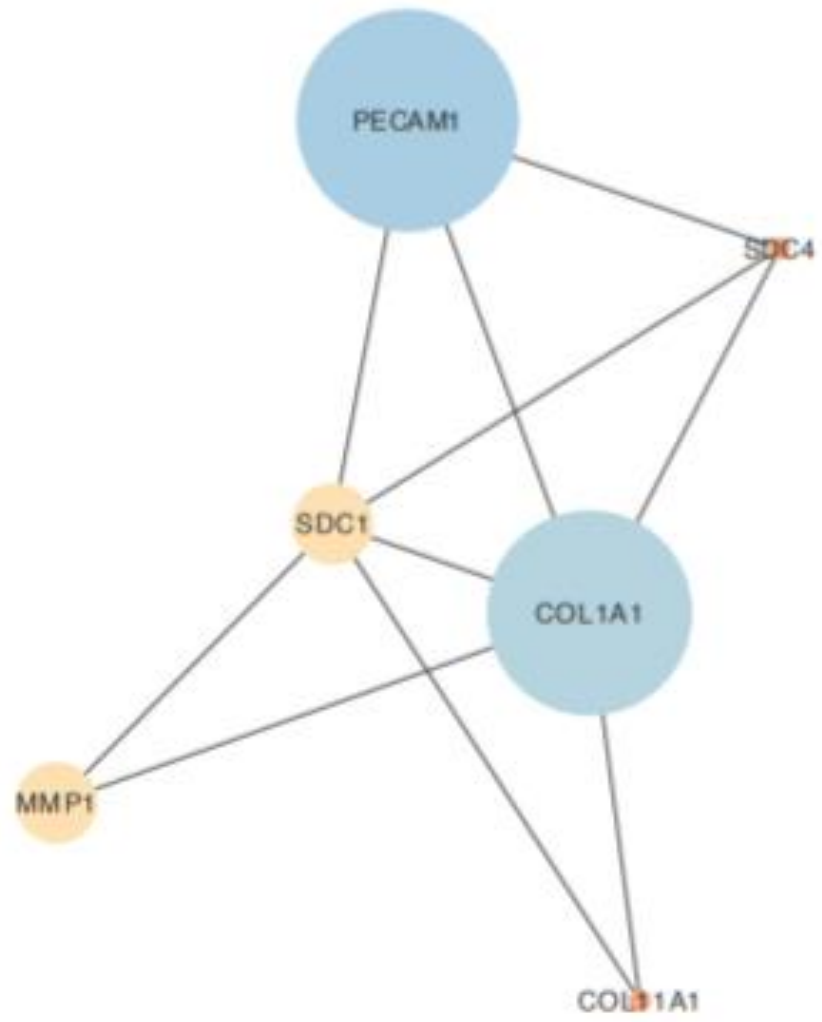

Cluster 3

MCODE score = 4.0

6 nodes 10 edges

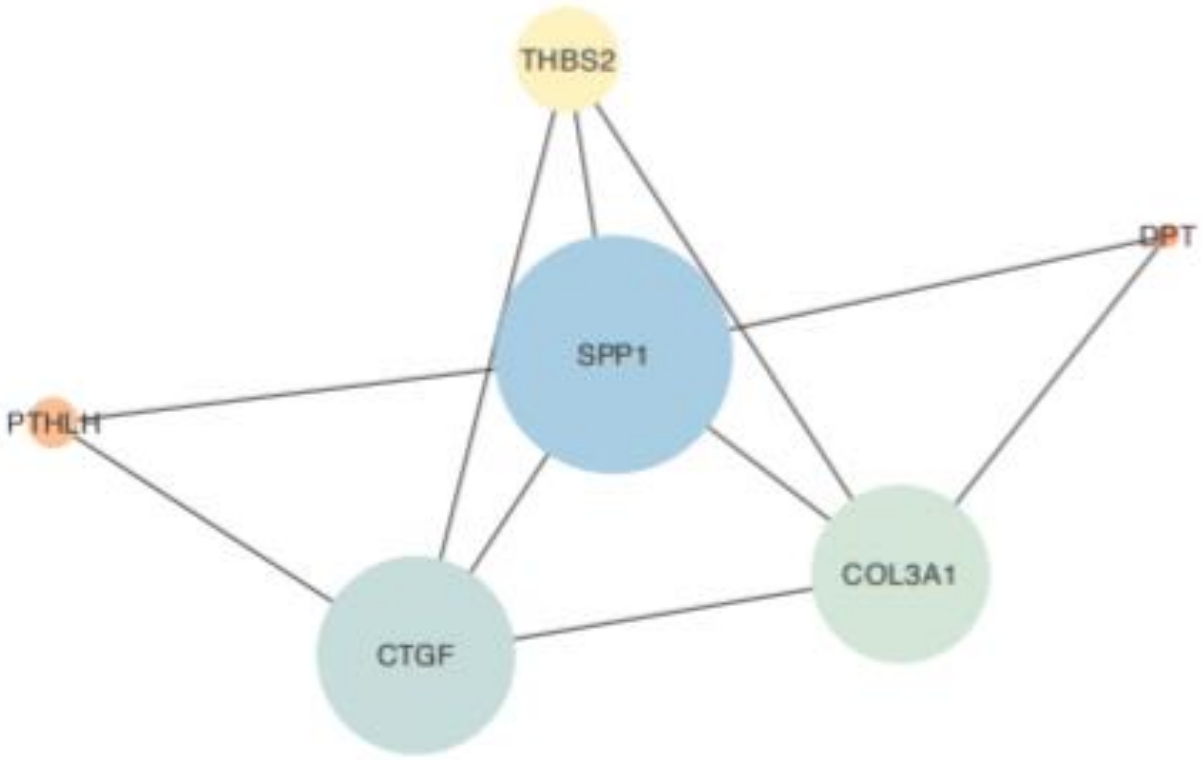

Cluster 4

MCODE score = 4.0

6 nodes 10 edges

Figure 1. The other three cluster in PPI network.

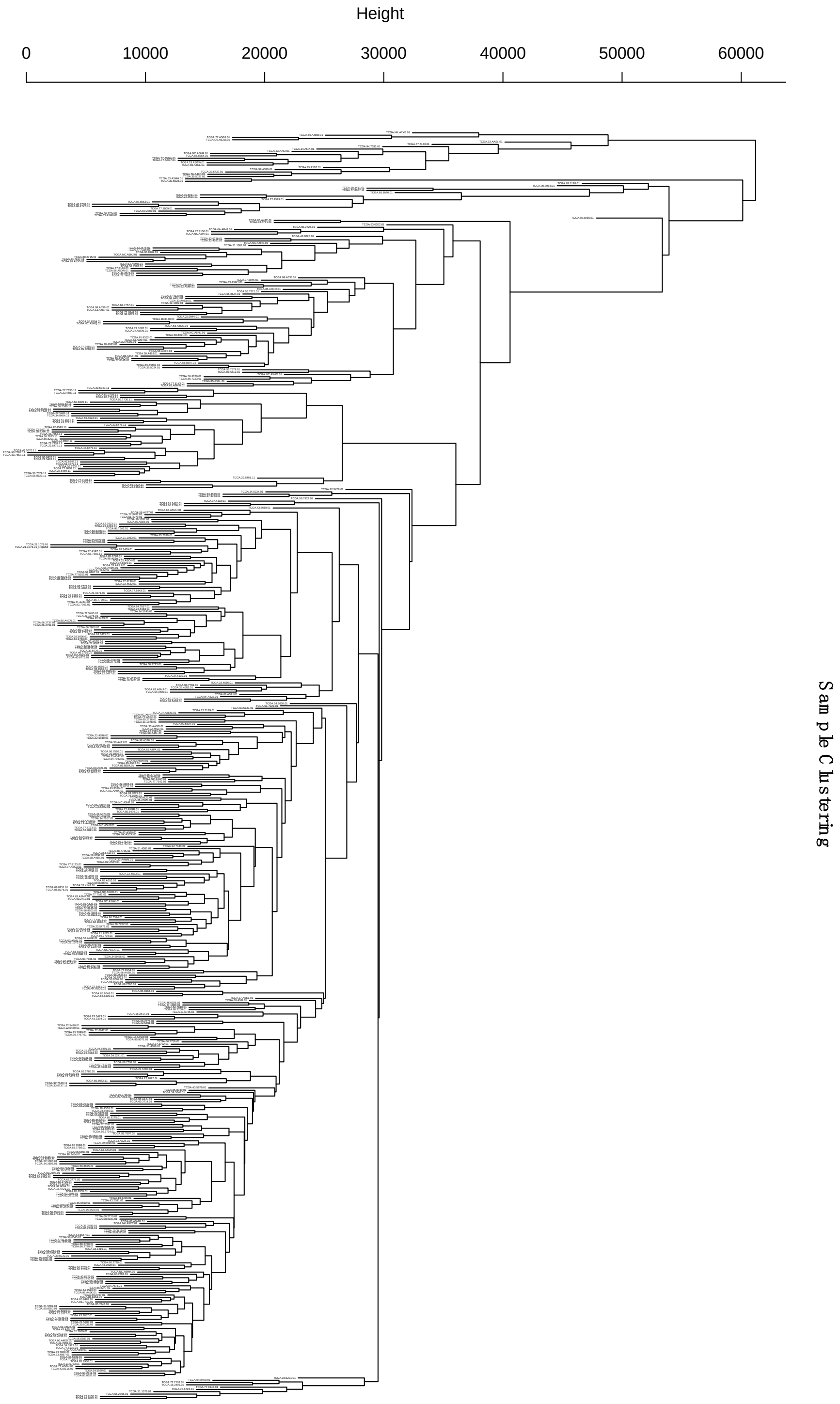

Figure 2. A tree diagram of the hierarchical clustering results before the sample is removed.

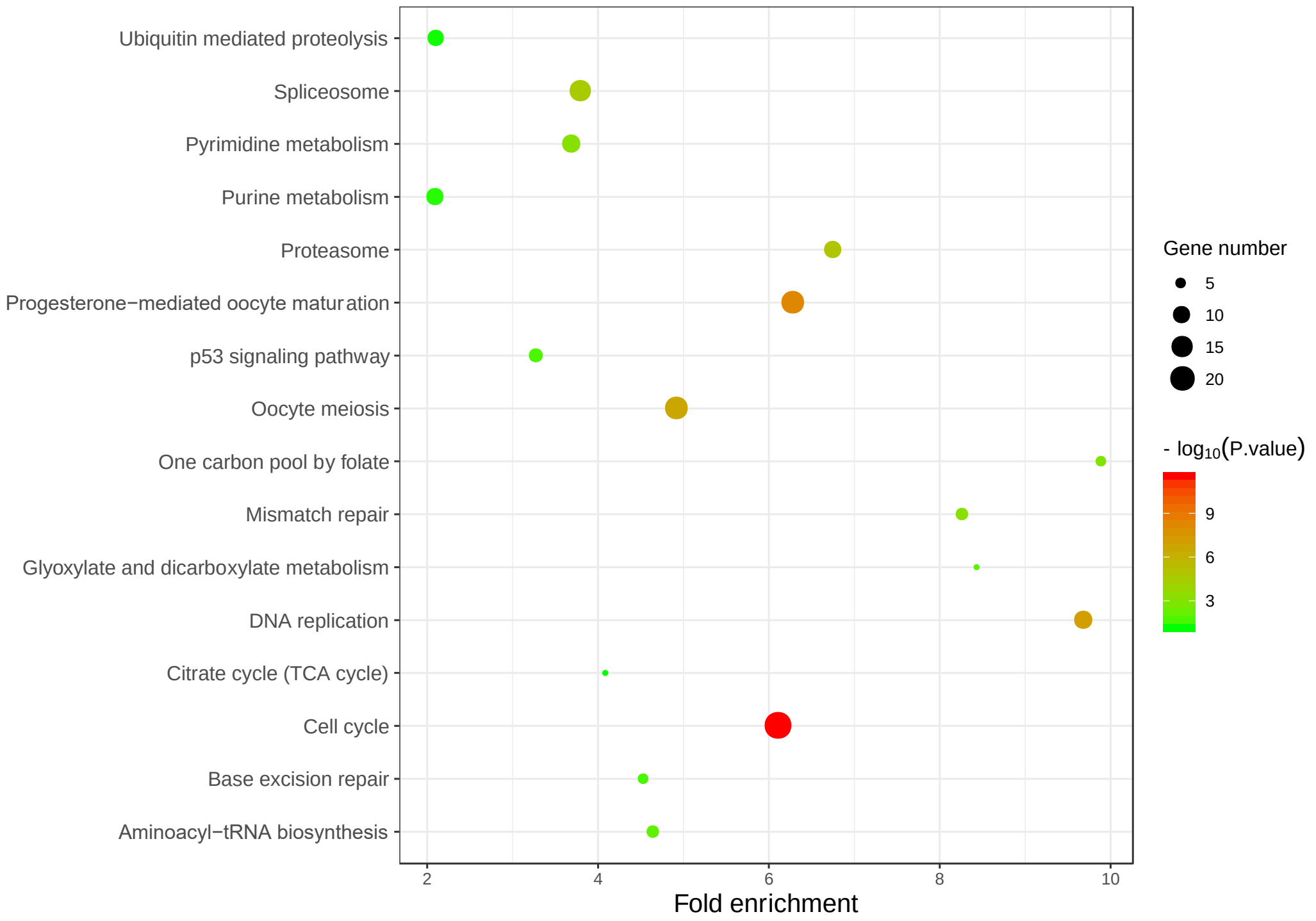

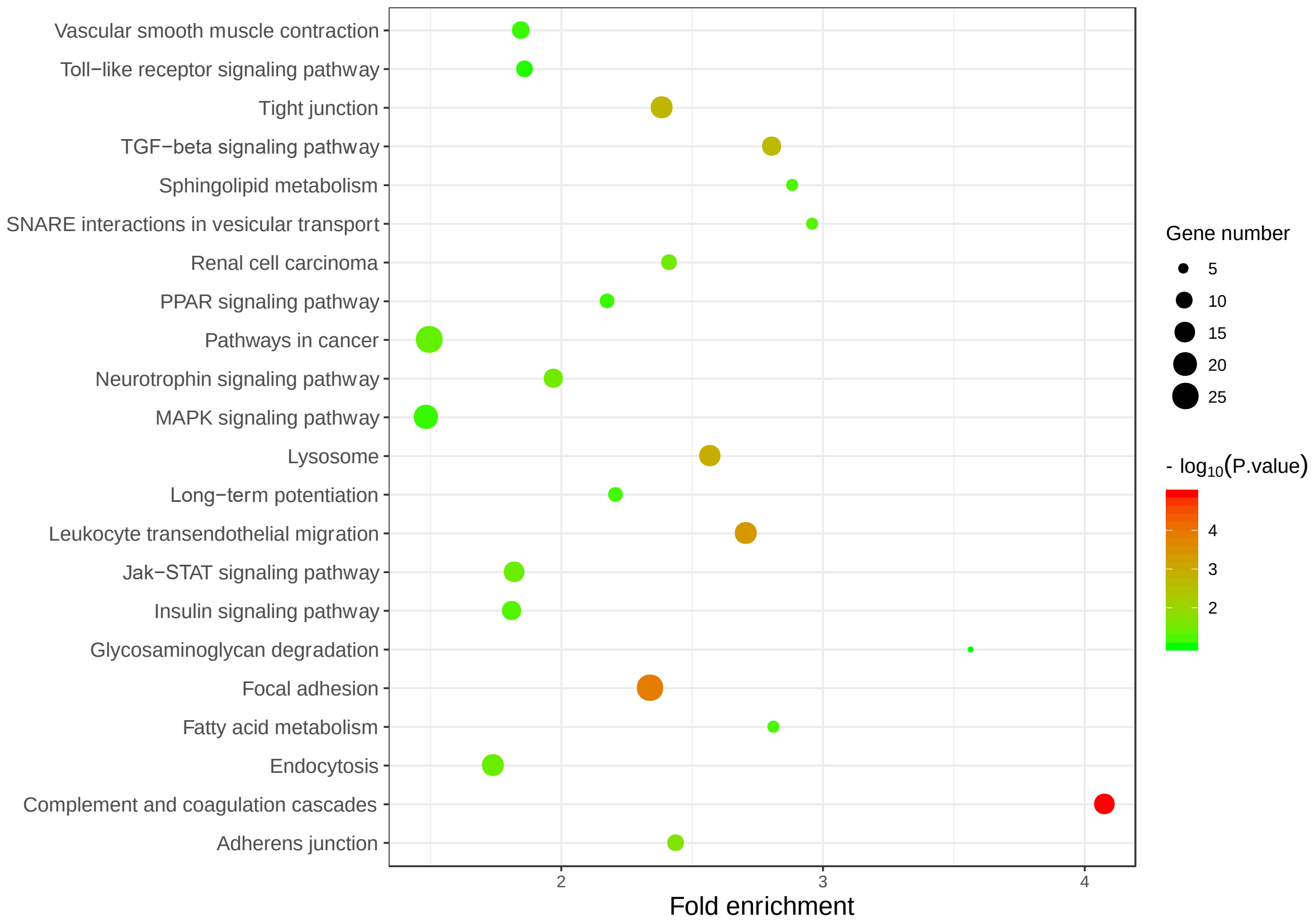

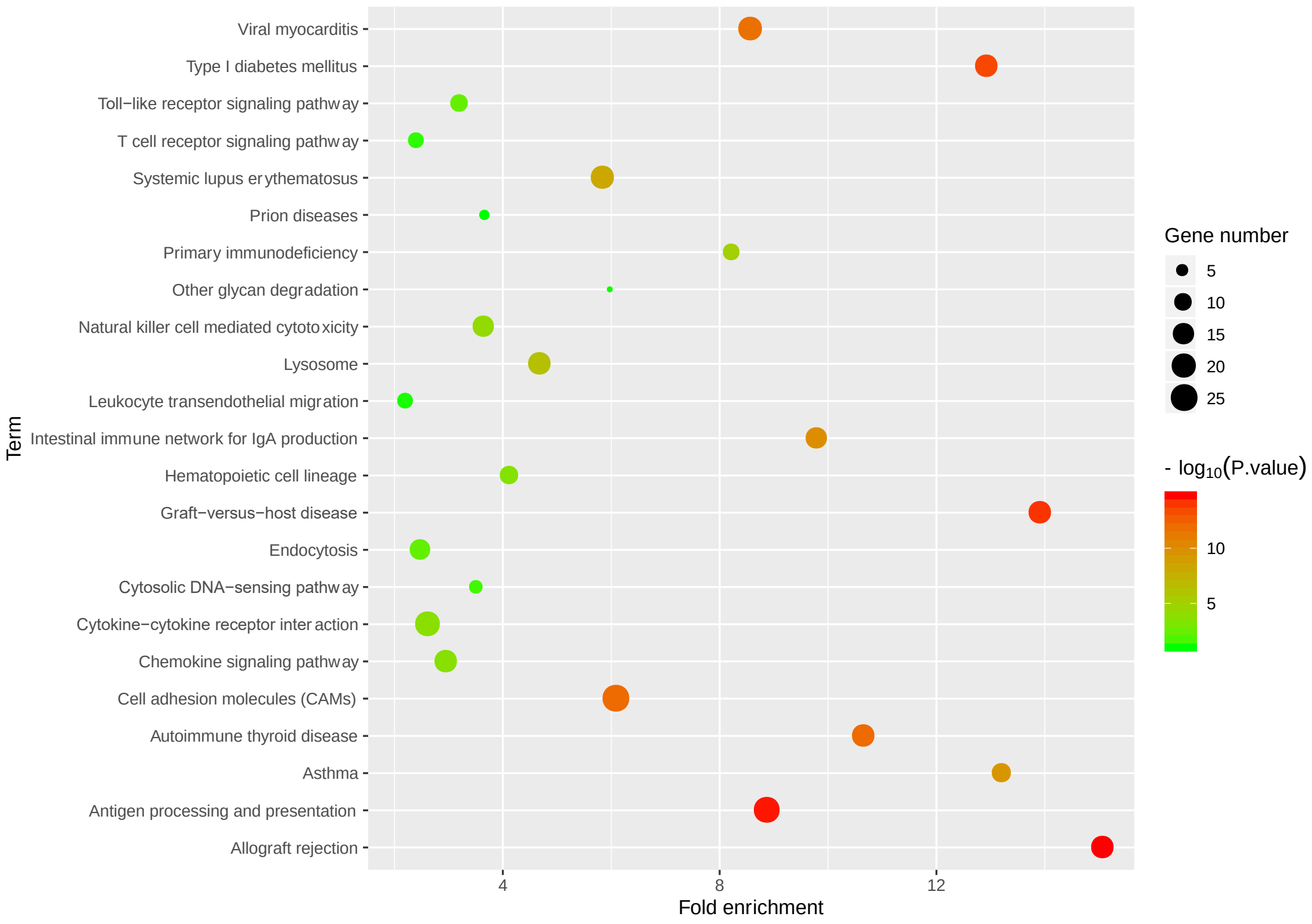

Figure3. The bubble figure of the three gene modules KEGG pathway analysis. Top, middle and bottom are the results of blue, turquoise and yellow respectively. The bubble size is the number of genes, and the bubble color represents the magnitude of the significance. The abscissa is the degree of enrichment, and the ordinate is the different regulatory pathway items.

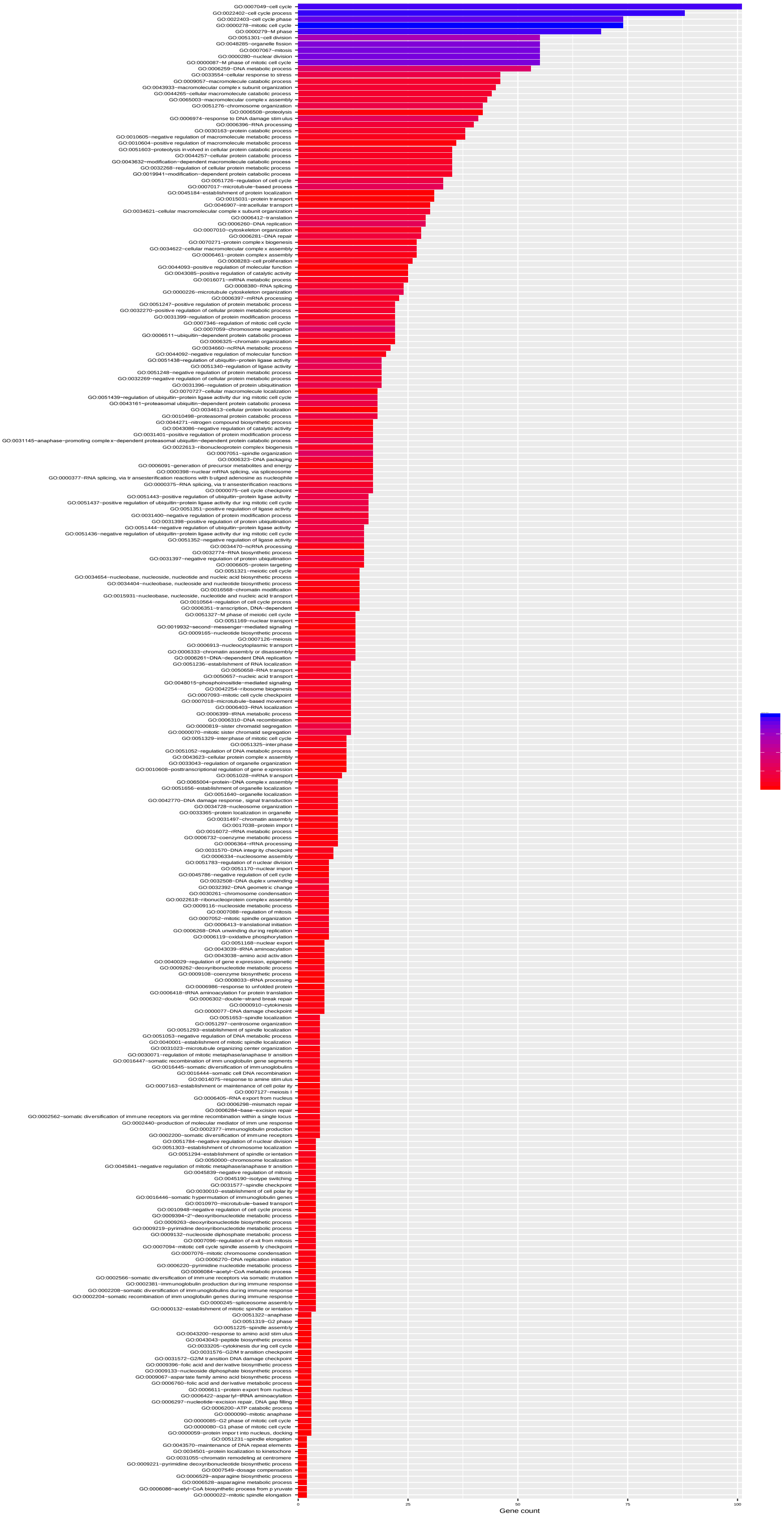
GO:0007049~cell cycle
GO:0022402~cell cycle process
GO:0022403~cell cycle phase
GO:0000278~mitotic cell cycle
GO:0000279~M phase
GO:0051301~cell division
GO:0048285~organelle fission
GO:0007067~mitosis
GO:0000280~nuclear division
GO:0000087~M phase of mitotic cell cycle
GO:0006259~DNA metabolic process
GO:0033554~cellular response to stress
GO:0009057~macromolecule catabolic process
GO:0043933~macromolecular complex subunit organization
GO:0044265~cellular macromolecule catabolic process
GO:0065003~macromolecular complex assembly
GO:0051276~chromosome organization
GO:0006508~proteolysis
GO:0006974~response to DNA damage stimulus
GO:0006396~RNA processing
GO:0030163~protein catabolic process
GO:0010605~negative regulation of macromolecule metabolic process
GO:0010604~positive regulation of macromolecule metabolic process
GO:0051603~proteolysis involved in cellular protein catabolic process
GO:0044257~cellular protein catabolic process
GO:0043632~modification-dependent macromolecule catabolic process
GO:0032268~regulation of cellular protein metabolic process
GO:0019941~modification-dependent protein catabolic process
GO:0051726~regulation of cell cycle
GO:0007017~microtubule-based process
GO:0045184~establishment of protein localization
GO:0015031~protein transport
GO:0046907~intracellular transport
GO:0034621~cellular macromolecular complex subunit organization
GO:0006412~translation
GO:0006260~DNA replication
GO:0007010~cytoskeleton organization
GO:0006281~DNA repair
GO:0070271~protein complex biogenesis
GO:0034622~cellular macromolecular complex assembly
GO:0006461~protein complex assembly
GO:0008283~cell proliferation
GO:0044093~positive regulation of molecular function
GO:0043085~positive regulation of catalytic activity
GO:0016071~mRNA metabolic process
GO:0008380~RNA splicing
GO:0000226~microtubule cytoskeleton organization
GO:0006397~mRNA processing
GO:0051247~positive regulation of protein metabolic process
GO:0032270~positive regulation of cellular protein metabolic process
GO:0031399~regulation of protein modification process
GO:0007346~regulation of mitotic cell cycle
GO:0007059~chromosome segregation
GO:0006511~ubiquitin-dependent protein catabolic process
GO:0006325~chromatin organization
GO:0034660~ncRNA metabolic process
GO:0044092~negative regulation of molecular function
GO:0051438~regulation of ubiquitin-protein ligase activity
GO:0051340~regulation of ligase activity
GO:0051248~negative regulation of protein metabolic process
GO:0032269~negative regulation of cellular protein metabolic process
GO:0031396~regulation of protein ubiquitination
GO:0070727~cellular macromolecule localization
GO:0051439~regulation of ubiquitin-protein ligase activity during mitotic cell cycle
GO:0043161~proteasomal ubiquitin-dependent protein catabolic process
GO:0034613~cellular protein localization
GO:0010498~proteasomal protein catabolic process
GO:0044271~nitrogen compound biosynthetic process
GO:0043086~negative regulation of catalytic activity
GO:0031401~positive regulation of protein modification process
GO:0031145~anaphase-promoting complex-dependent proteasomal ubiquitin-dependent protein catabolic process
GO:0022613~ribonucleoprotein complex biogenesis
GO:0007051~spindle organization
GO:0006323~DNA packaging
GO:0006091~generation of precursor metabolites and energy
GO:0000398~nuclear mRNA splicing, via spliceosome
GO:0000377~RNA splicing, via transesterification reactions with bulged adenosine as nucleophile
GO:0000375~RNA splicing, via transesterification reactions
GO:0000075~cell cycle checkpoint
GO:0051443~positive regulation of ubiquitin-protein ligase activity
GO:0051437~positive regulation of ubiquitin-protein ligase activity during mitotic cell cycle
GO:0051351~positive regulation of ligase activity
GO:0031400~negative regulation of protein modification process
GO:0031398~positive regulation of protein ubiquitination
GO:0051444~negative regulation of ubiquitin-protein ligase activity
GO:0051436~negative regulation of ubiquitin-protein ligase activity during mitotic cell cycle
GO:0051352~negative regulation of ligase activity
GO:0034470~ncRNA processing
GO:0032774~RNA biosynthetic process
GO:0031397~negative regulation of protein ubiquitination
GO:0006605~protein targeting
GO:0051321~meiotic cell cycle
GO:0034654~nucleobase, nucleoside, nucleotide and nucleic acid biosynthetic process
GO:0034404~nucleobase, nucleoside and nucleotide biosynthetic process
GO:0016568~chromatin modification
GO:0015931~nucleobase, nucleoside, nucleotide and nucleic acid transport
GO:0010564~regulation of cell cycle process
GO:0006351~transcription, DNA-dependent
GO:0051327~M phase of meiotic cell cycle
GO:0051169~nuclear transport
GO:0019932~second-messenger-mediated signaling
GO:0009165~nucleotide biosynthetic process
GO:0007126~meiosis
GO:0006913~nucleocytoplasmic transport
GO:0006333~chromatin assembly or disassembly
GO:0006261~DNA-dependent DNA replication
GO:0051236~establishment of RNA localization
GO:0050658~RNA transport
GO:0050657~nucleic acid transport
GO:0048015~phosphoinositide-mediated signaling
GO:0042254~ribosome biogenesis
GO:0007093~mitotic cell cycle checkpoint
GO:0007018~microtubule-based movement
GO:0006403~RNA localization
GO:0006399~tRNA metabolic process
GO:0006310~DNA recombination
GO:0000819~sister chromatid segregation
GO:0000070~mitotic sister chromatid segregation
GO:0051329~interphase of mitotic cell cycle
GO:0051325~interphase
GO:0051052~regulation of DNA metabolic process
GO:0043623~cellular protein complex assembly
GO:0033043~regulation of organelle organization
GO:0010608~posttranscriptional regulation of gene expression
GO:0051028~mRNA transport
GO:0065004~protein-DNA complex assembly
GO:0051656~establishment of organelle localization
GO:0051640~organelle localization
GO:0042770~DNA damage response, signal transduction
GO:0034728~nucleosome organization
GO:0033365~protein localization in organelle
GO:0031497~chromatin assembly
GO:0017038~protein import
GO:0016072~rRNA metabolic process
GO:0006732~coenzyme metabolic process
GO:0006364~rRNA processing
GO:0031570~DNA integrity checkpoint
GO:0006334~nucleosome assembly
GO:0051783~regulation of nuclear division
GO:0051170~nuclear import
GO:0045786~negative regulation of cell cycle
GO:0032508~DNA duplex unwinding
GO:0032392~DNA geometric change
GO:0030261~chromosome condensation
GO:0022618~ribonucleoprotein complex assembly
GO:0009116~nucleoside metabolic process
GO:0007088~regulation of mitosis
GO:0007052~mitotic spindle organization
GO:0006413~translational initiation
GO:0006268~DNA unwinding during replication
GO:0006119~oxidative phosphorylation
GO:0051168~nuclear export
GO:0043039~tRNA aminoacylation
GO:0043038~amino acid activation
GO:0040029~regulation of gene expression, epigenetic
GO:0009262~deoxyribonucleotide metabolic process
GO:0009108~coenzyme biosynthetic process
GO:0008033~tRNA processing
GO:0006986~response to unfolded protein
GO:0006418~tRNA aminoacylation for protein translation
GO:0006302~double-strand break repair
GO:0000910~cytokinesis
GO:0000077~DNA damage checkpoint
GO:0051653~spindle localization
GO:0051297~centrosome organization
GO:0051293~establishment of spindle localization
GO:0051053~negative regulation of DNA metabolic process
GO:0040001~establishment of mitotic spindle localization
GO:0031023~microtubule organizing center organization
GO:0030071~regulation of mitotic metaphase/anaphase transition
GO:0016447~somatic recombination of immunoglobulin gene segments
GO:0016445~somatic diversification of immunoglobulins
GO:0016444~somatic cell DNA recombination
GO:0014075~response to amine stimulus
GO:0007163~establishment or maintenance of cell polarity
GO:0007127~meiosis I
GO:0006405~RNA export from nucleus
GO:0006298~mismatch repair
GO:0006284~base-excision repair
GO:0002562~somatic diversification of immune receptors via germline recombination within a single locus
GO:0002440~production of molecular mediator of immune response
GO:0002377~immunoglobulin production
GO:0002200~somatic diversification of immune receptors
GO:0051784~negative regulation of nuclear division
GO:0051303~establishment of chromosome localization
GO:0051294~establishment of spindle orientation
GO:0050000~chromosome localization
GO:0045841~negative regulation of mitotic metaphase/anaphase transition
GO:0045839~negative regulation of mitosis
GO:0045190~isotype switching
GO:0031577~spindle checkpoint
GO:0030010~establishment of cell polarity
GO:0016446~somatic hypermutation of immunoglobulin genes
GO:0010970~microtubule-based transport
GO:0010948~negative regulation of cell cycle process
GO:0009394~2'-deoxyribonucleotide metabolic process
GO:0009263~deoxyribonucleotide biosynthetic process
GO:0009219~pyrimidine deoxyribonucleotide metabolic process
GO:0009132~nucleoside diphosphate metabolic process
GO:0007096~regulation of exit from mitosis
GO:0007094~mitotic cell cycle spindle assembly checkpoint
GO:0007076~mitotic chromosome condensation
GO:0006270~DNA replication initiation
GO:0006220~pyrimidine nucleotide metabolic process
GO:0006084~acetyl-CoA metabolic process
GO:0002566~somatic diversification of immune receptors via somatic mutation
GO:0002381~immunoglobulin production during immune response
GO:0002208~somatic diversification of immunoglobulins during immune response
GO:0002204~somatic recombination of immunoglobulin genes during immune response
GO:0000245~spliceosome assembly
GO:0000132~establishment of mitotic spindle orientation
GO:0051322~anaphase
GO:0051319~G2 phase
GO:0051225~spindle assembly
GO:0043200~response to amino acid stimulus
GO:0043043~peptide biosynthetic process
GO:0033205~cytokinesis during cell cycle
GO:0031576~G2/M transition checkpoint
GO:0031572~G2/M transition DNA damage checkpoint
GO:0009396~folic acid and derivative biosynthetic process
GO:0009133~nucleoside diphosphate biosynthetic process
GO:0009067~aspartate family amino acid biosynthetic process
GO:0006760~folic acid and derivative metabolic process
GO:0006611~protein export from nucleus
GO:0006422~aspartyl-tRNA aminoacylation
GO:0006297~nucleotide-excision repair, DNA gap filling
GO:0006200~ATP catabolic process
GO:0000090~mitotic anaphase
GO:0000085~G2 phase of mitotic cell cycle
GO:0000080~G1 phase of mitotic cell cycle
GO:0000059~protein import into nucleus, docking
GO:0051231~spindle elongation
GO:0043570~maintenance of DNA repeat elements
GO:0034501~protein localization to kinetochore
GO:0031055~chromatin remodeling at centromere
GO:0009221~pyrimidine deoxyribonucleotide biosynthetic process
GO:0007549~dosage compensation
GO:0006529~asparagine biosynthetic process
GO:0006528~asparagine metabolic process
GO:0006086~acetyl-CoA biosynthetic process from pyruvate
GO:0000022~mitotic spindle elongation
0
25
50
75
100
Gene count

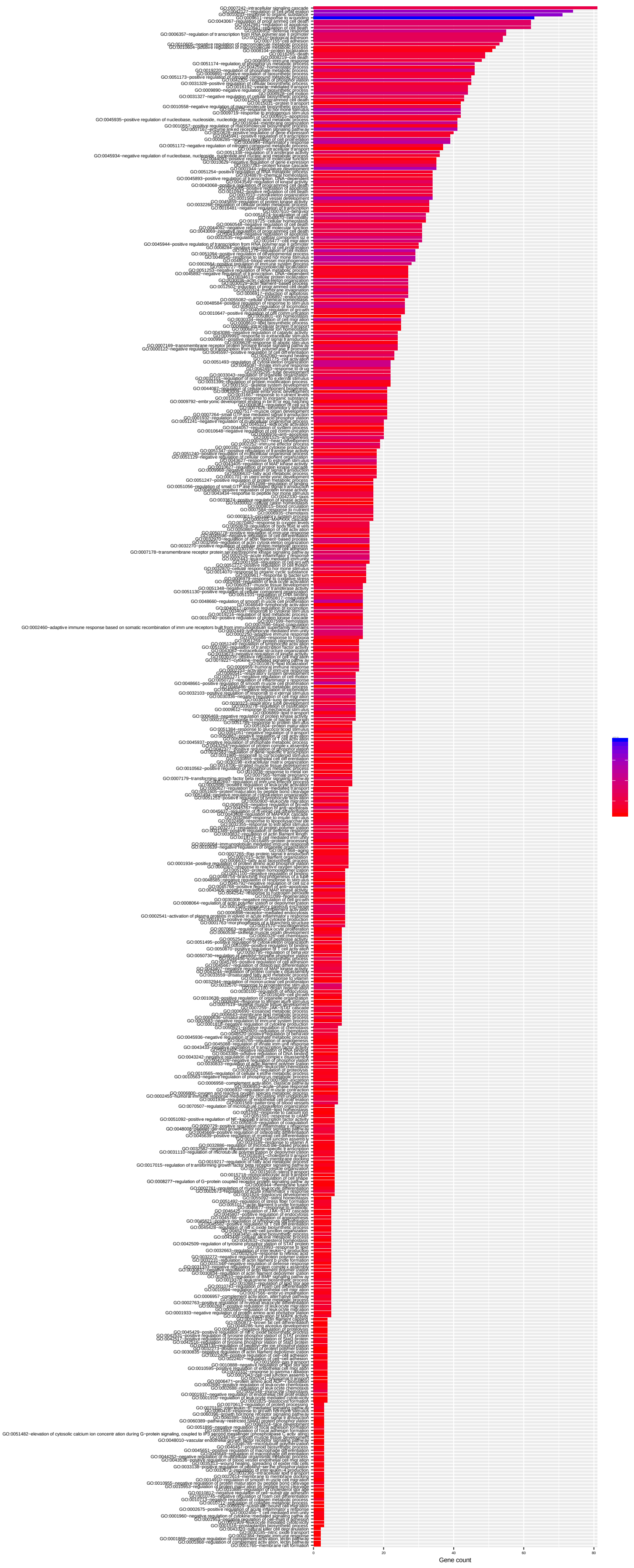

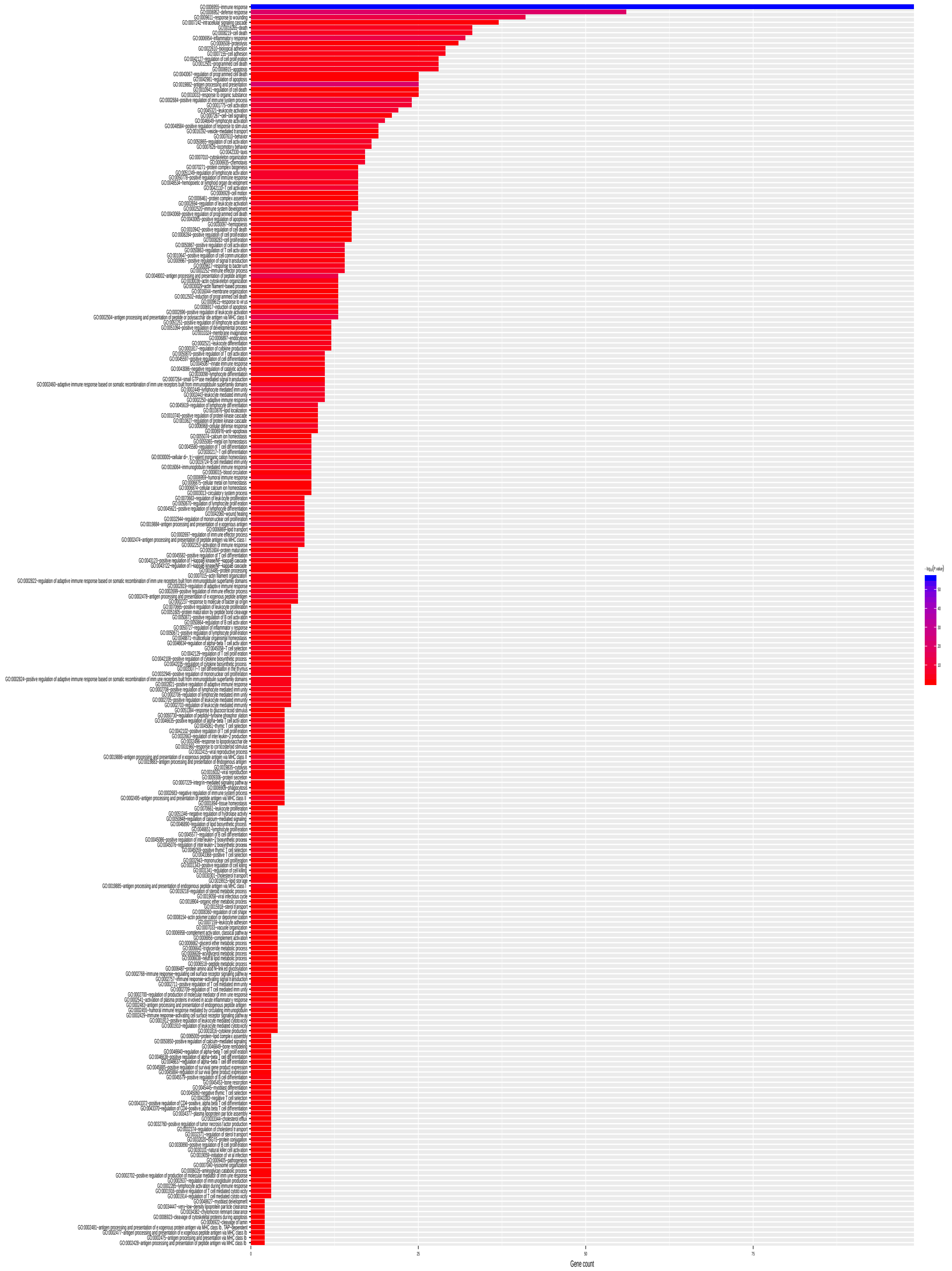

Figure 4. The figure of the three gene modules GO analysis. Top, middle and bottom are the results of blue, turquoise and yellow respectively. The y-axis shows significantly enriched GO terms, and x-axis represents significance of enrichment.